\documentclass{apj}
\usepackage{CJKutf8}
\usepackage[caption=false]{subfig}
\usepackage{graphicx}
\usepackage{amsmath}
\usepackage{extarrows}
\usepackage{multirow}
\usepackage{graphicx}
\usepackage{booktabs}
\usepackage{amsmath,amssymb}
\usepackage[T1]{fontenc}
\graphicspath{{Pic/}}
\usepackage{longtable}

\usepackage{threeparttablex}
\begin{document}

\title{Wide-Field  X-ray Polarimetry for High Energy Astronomical Transients: First results of the pathfinder CXPD Cubesat Mission}

\correspondingauthor{Hong-Bang Liu, Yang-Heng Zheng, En-Wei Liang}
\email{E-mail: liuhb@gxu.edu.cn; zhengyh@ucas.ac.cn; lew@gxu.edu.cn}

\author{Hong-Bang Liu}
\affiliation{Guangxi Key Laboratory for Relativistic Astrophysics, School of Physical Science and Technology, Guangxi University, Nanning 530004, People's Republic of China}

\author{Zu-Ke Feng}
\affiliation{Guangxi Key Laboratory for Relativistic Astrophysics, School of Physical Science and Technology, Guangxi University, Nanning 530004, People's Republic of China}

\author{Huan-Bo Feng}
\affiliation{Guangxi Key Laboratory for Relativistic Astrophysics, School of Physical Science and Technology, Guangxi University, Nanning 530004, People's Republic of China}

\author{Di-Fan Yi}
\affiliation{School of Physics Science, University of Chinese Academy of Sciences, Beijing 100049, People's Republic of China}

\author{Li-Rong Xie}
\affiliation{Guangxi Key Laboratory for Relativistic Astrophysics, School of Physical Science and Technology, Guangxi University, Nanning 530004, People's Republic of China}

\author{Yan-Jun Xie}
\affiliation{Guangxi Key Laboratory for Relativistic Astrophysics, School of Physical Science and Technology, Guangxi University, Nanning 530004, People's Republic of China}

\author{Zong-Wang Fan}
\affiliation{Guangxi Key Laboratory for Relativistic Astrophysics, School of Physical Science and Technology, Guangxi University, Nanning 530004, People's Republic of China}

\author{Jin Zhang}
\affiliation{Guangxi Key Laboratory for Relativistic Astrophysics, School of Physical Science and Technology, Guangxi University, Nanning 530004, People's Republic of China}

\author{Wen-Jin Xie}
\affiliation{Guangxi Key Laboratory for Relativistic Astrophysics, School of Physical Science and Technology, Guangxi University, Nanning 530004, People's Republic of China}

\author{Xue-Feng Huang}
\affiliation{Guangxi Key Laboratory for Relativistic Astrophysics, School of Physical Science and Technology, Guangxi University, Nanning 530004, People's Republic of China}

\author{Wei Deng}
\affiliation{Guangxi Key Laboratory for Relativistic Astrophysics, School of Physical Science and Technology, Guangxi University, Nanning 530004, People's Republic of China}

\author{Fei Xie}
\affiliation{Guangxi Key Laboratory for Relativistic Astrophysics, School of Physical Science and Technology, Guangxi University, Nanning 530004, People's Republic of China}

\author{Dong Wang}
\affiliation{PLAC, Key Laboratory of Quark \& Lepton Physics (MOE), Central China Normal University, Wuhan 430079, People's Republic of China}

\author{Zi-Li Li}
\affiliation{PLAC, Key Laboratory of Quark \& Lepton Physics (MOE), Central China Normal University, Wuhan 430079, People's Republic of China}

\author{Hui Wang}
\affiliation{PLAC, Key Laboratory of Quark \& Lepton Physics (MOE), Central China Normal University, Wuhan 430079, People's Republic of China}

\author{Ran Chen}
\affiliation{PLAC, Key Laboratory of Quark \& Lepton Physics (MOE), Central China Normal University, Wuhan 430079, People's Republic of China}

\author{Shi-Qiang Zhou}
\affiliation{PLAC, Key Laboratory of Quark \& Lepton Physics (MOE), Central China Normal University, Wuhan 430079, People's Republic of China}

\author{Kai Chen}
\affiliation{PLAC, Key Laboratory of Quark \& Lepton Physics (MOE), Central China Normal University, Wuhan 430079, People's Republic of China}

\author{Jin Li}
\affiliation{Key Laboratory of Particle Astrophysics, Institute of High Energy Physics, Chinese Academy of Sciences, Beijing 100049, People's Republic of China}
\affiliation{School of Physics Science, University of Chinese Academy of Sciences, Beijing 100049, People's Republic of China}

\author{Qian Liu}
\affiliation{School of Physics Science, University of Chinese Academy of Sciences, Beijing 100049, People's Republic of China}

\author{Shi Chen}
\affiliation{School of Physics Science, University of Chinese Academy of Sciences, Beijing 100049, People's Republic of China}

\author{Rui-Ting Ma}
\affiliation{School of Physics Science, University of Chinese Academy of Sciences, Beijing 100049, People's Republic of China}

\author{Bin-Long Wang}
\affiliation{School of Physics Science, University of Chinese Academy of Sciences, Beijing 100049, People's Republic of China}

\author{Zhen-Yu Tang}
\affiliation{Beijing Institute of Spacecraft Environment Engineering, Beijing 100094, People's Republic of China}

\author{Hang-Zhou Li}
\affiliation{The 13th Research Institute, China Electronics Technology Group Corporation (CETC), Shijiazhuang 050051, People's Republic of China}

\author{Bo Peng}
\affiliation{The 13th Research Institute, China Electronics Technology Group Corporation (CETC), Shijiazhuang 050051, People's Republic of China}

\author{Shu-Lin Liu}
\affiliation{Key Laboratory of Particle Astrophysics, Institute of High Energy Physics, Chinese Academy of Sciences, Beijing 100049, People's Republic of China}

\author{Xiang-Ming Sun}
\affiliation{PLAC, Key Laboratory of Quark \& Lepton Physics (MOE), Central China Normal University, Wuhan 430079, People's Republic of China}

\author{Yang-Heng Zheng}
\affiliation{School of Physics Science, University of Chinese Academy of Sciences, Beijing 100049, People's Republic of China}

\author{En-Wei Liang}
\affiliation{Guangxi Key Laboratory for Relativistic Astrophysics, School of Physical Science and Technology, Guangxi University, Nanning 530004, People's Republic of China}

% \footnotetext[1]{These authors contributed equally to this work.}
% \thanks{These authors contributed equally.}

\begin{abstract}
The Low Energy Polarization Detector (LPD) is a key component of the next-generation large-scale Gamma-Ray Burst polarimeter, POLAR-2. It is designed for polarization observations of transient sources in the soft X-ray energy range with a wide field of view (FOV). To validate the key technologies required for wide-FOV X-ray polarization measurements, the Cosmic X-ray Polarization Detector (CXPD) CubeSat was developed as a prototype for the LPD.
The CXPD is equipped with two Gas Microchannel Plate Pixel Detectors (GMPDs) that measure X-ray polarization via the photoelectric effect, where ejected photoelectrons produce ionization tracks in the gas which are imaged to reconstruct their emission directions. Laboratory calibrations of the modulation factor and energy spectra were successfully performed using linear polarized X-ray sources at 2.98 keV, 4.51 keV, 6.40 keV, and 8.05 keV. 
Since its launch in June 2023, the CXPD has successfully completed critical in-orbit technology verification. It has also performed polarization observations of two bright X-ray sources—Sco X-1 and the transient Swift J1727.8-1613—yielding constraints on their polarization degrees and angles. Notably, this was the first time that an anti-coincidence detector had been implemented in an X-ray polarimeter, enabling in-orbit verification of the charged-particle background rejection algorithm. These results demonstrate the feasibility of wide-field soft X-ray polarization measurements and provide essential guidance for the development of the LPD for the POLAR-2 mission, thereby advancing the frontier of X-ray polarization astronomy.
\end{abstract}


% \maketitle

%________________________________________________ sections below
%
\section{Introduction}
\label{section1}
\begin{figure*}
\centering
\includegraphics[scale=0.5]{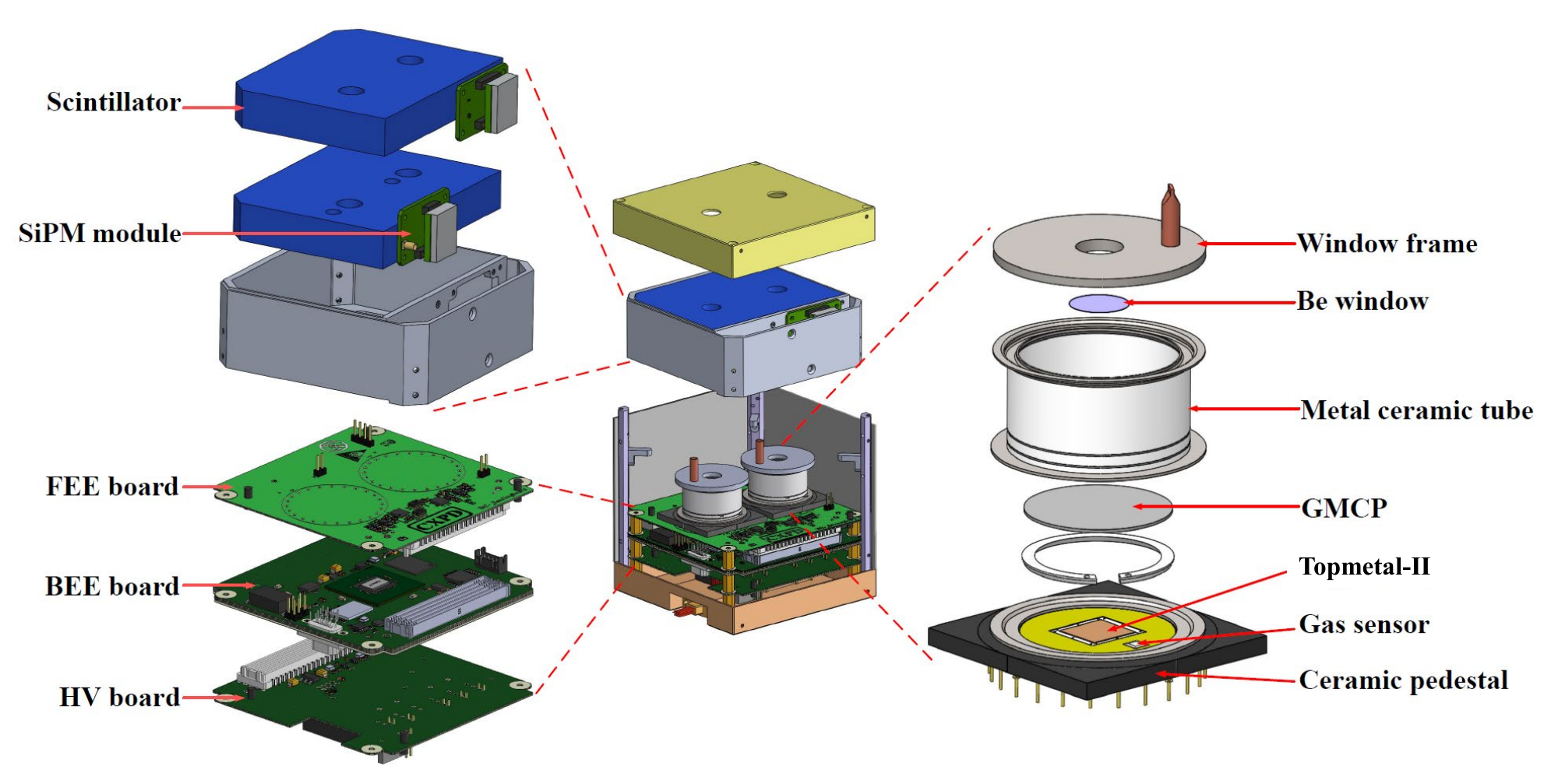}
\caption{Structure of the CXPD CubeSat and GMPD unit.}
\label{cxpd_structure}
\end{figure*}
Gamma-ray bursts (GRBs) are among the most energetic events in the universe, releasing enormous amounts of energy within seconds. Despite extensive observations, the physical mechanisms driving GRBs—including jet composition, magnetic field structures, and radiation processes—are still not well understood \cite{galaxies9040082}. Polarization offers a new observational dimension, and combining polarization measurements with spectral data can provide deeper insights into these unresolved questions \cite{kumar2015physics,10.1111/j.1745-3933.2009.00624.x,10.1093/mnras/sts219,10.1111/j.1365-2966.2010.17600.x,Zhang_2002,Bégué_2016,Rees_1994}. These theoretical studies mainly focus on the prompt emission, which originates directly from the relativistic jet and carries key information about the central engine. As the earliest and brightest phase of a GRB, the prompt emission is predicted to exhibit relatively high polarization degrees, reflecting the magnetic field geometry and radiation mechanisms in the innermost emission region \cite{Toma_2009}.

% \Authorfootnote

% \noindent  

Recent advances in polarimetry have yielded intriguing results: the POLAR mission detected low average polarization degrees (PDs) of approximately 10\% for GRB prompt emissions \cite{zhang2019detailed}. This relatively low PD may also be partially attributed to the evolution of the polarization angle (PA) within individual pulses. These findings challenge traditional synchrotron models that predict higher PDs \cite{granot2003linear} and contrast with previous observations reporting high PDs for GRBs \cite{guan2023interpreting}. Furthermore, IXPE observed GRB 221009A and provided the first X-ray polarization constraints for both the prompt and afterglow phases, setting 99\% confidence upper limits of 55–82\% and 13.8\%, respectively \cite{negro2023ixpe}. These results indicate that the afterglow emission exhibits low polarization, consistent with a disordered magnetic field in the external shock, while the prompt phase could still allow for higher intrinsic polarization. This discrepancy underscores the complexity of GRB emission mechanisms and highlights the need for enhanced observational capabilities and more advanced theoretical frameworks.

POLAR-2 is being developed as the successor to POLAR, aiming to further investigate the polarization properties of GRB prompt emission with improved sensitivity and an extended energy coverage. Theoretical studies indicate that different emission models predict varying polarization degrees depending on the observed energy range. For example, synchrotron radiation and Compton drag models predict lower polarization degrees at lower energies \cite{toma2009statistical}, whereas the photosphere model suggests more prominent polarization at several keV \cite{lundman2018polarization}.

To explore the largely uncharted soft X-ray polarization regime, POLAR-2 is equipped with a Low-Energy X-ray Polarization Detector (LPD) \cite{feng2023orbit}, capable of measuring polarization in the 2–10 keV range using an array of X-ray photoelectric polarimeters. Unlike missions such as IXPE and eXTP \cite{weisskopf2022imaging, zhang2019enhanced}, which rely on focusing optics for long-term observations of persistent sources. As mentioned above, IXPE’s constraints on GRB prompt emission were not obtained via direct detection of prompt radiation but were instead inferred from the polarization of dust-scattered X-ray halos. By contrast, the LPD adopts a wide field-of-view (FOV) design optimized for sky surveys of transient sources.

To validate the key technologies required for wide-FOV X-ray polarization measurements, the Cosmic X-ray Polarization Detector (CXPD) CubeSat was developed as a prototype for the LPD. Launched into a sun-synchronous orbit on June 7, 2023, CXPD serves as a testbed, paving the way for future polarization observations with POLAR-2/LPD.

% 引入polarlight

This paper introduces the design, ground calibration, and in-orbit operations and observation results of the CXPD. The remainder of the paper is organized as follows: Section 2 describes the detector structure and electronic systems; Section 3 presents performance and ground calibration results; Section 4 discusses in-orbit observations; and Section 5 provides the discussion and conclusion.

\section{Detector Structure}\label{section2}
The CXPD CubeSat's payload consists of the anti-coincidence detector, the Gas Microchannel Plate Pixel Detector (GMPD) \cite{feng2024gas}, the electronic systems, and the mechanical support structure. The structural design of these components is illustrated in Fig. \ref{cxpd_structure}. The anti-coincidence detector is positioned at the top of the payload and consists of two plastic scintillator detectors arranged in a top-down configuration. Its primary purpose is to provide timestamps for in-orbit verification of the track discrimination algorithm. Below the anti-coincidence detector, two GMPDs are fixed onto the first circuit board of the electronic system. The electronic system comprises three printed circuit boards arranged from top to bottom: the front-end electronics (FEE) board, the back-end electronics (BEE) board, and the high-voltage (HV) board. The following subsections provide detailed specifications for some key components.

\subsection{anti-coincidence detector}
The scintillator detector consists of two plastic scintillator detectors. Each plastic scintillator detector is composed of a plastic scintillator and a Hamamatsu S13365-3050SA silicon photomultiplier (SiPM) module. The dimensions of the plastic scintillator are 8.5 $\times$ 7 $\times$ 1.5 cm³. The SiPM module includes an MPPC (Multi-Pixel Photon Counter), a signal amplification circuit, a high-voltage power supply circuit, and a temperature compensation circuit. Its photosensitive area measures 3 $\times$ 3 mm², and the signal output is in analog form.  

When incident particles traverse the scintillator, they lose energy and deposit it within the scintillator material, causing ionization excitation of atoms in the scintillator. Subsequently, the excited atoms de-excite, emitting scintillation photons. These scintillation photons enter the semiconductor material, generating charge carriers (electron-hole pairs) due to the photoelectric effect. Under the influence of an electric field, these charge carriers undergo avalanche multiplication. The signal is then amplified by the amplification circuit, resulting in the output of an analog signal.

\subsection{The Gas Microchannel plate Pixel Detector}
GMPD is a two-dimensional gas proportional counter with pixel readout, used for capturing the track image of photoelectrons emitted after X-ray absorption \cite{feng2023spectral}. Incoming X-rays penetrate a beryllium window and are absorbed by the working gas in the GMPD. Upon absorption, a photoelectron is emitted and initiates the ionization of gas molecules into ions and primary electrons. Under the influence of a parallel electric field, primary electrons drift toward the anode. As these electrons pass through the micro-holes in the gas microchannel plate (GMCP), where the electric field is strong, avalanche multiplication occurs, resulting in the production of secondary electrons. The number of electrons increases by several thousand times (gain factor). Next, these secondary electrons drift along the field lines. Some of the electrons terminate at the bottom electrode of the GMCP, while others continue to drift toward the anode, which is the Topmetal pixel chip. Subsequently, the induced charge on the pixel is amplified and shaped by the charge-sensitive preamplifier (CSA) within the chip, completing the entire process of event detection.

The key components of the GMPD is described as followed:

-Window: The GMPD X-ray window is made of beryllium with a thickness of 100 $\mu\mathrm{m}$ and is mounted beneath a supporting Kovar alloy frame. The frame has a circular opening of 10 mm in diameter, allowing X-rays to enter the detector. The beryllium window is brazed to the Kovar frame, ensuring gas-tightness and providing electrical contact to form the detector’s cathode. During operation, the cathode is biased at a high voltage of approximately 3200 V.

-GMCP: The GMCP is a microchannel plate (MCP) that can operate in a gas environment \cite{liu2011study, gys2015micro}. The central insulator is made of bismuth silicate glass with a thickness of 300 $\mu\mathrm{m}$, and the electrodes on both sides are composed of a Ni-Cr alloy with a penetration depth of 25–30 $\mu\mathrm{m}$. The microchannels have a diameter of 50 $\mu\mathrm{m}$ and a spacing of 60 $\mu\mathrm{m}$. After hydrogen reduction, the GMCP exhibits a bulk resistance on the order of $\mathrm{G\Omega}$ \cite{feng2023charging}. During operation, a bias voltage of 950–1050 V is applied between the upper and lower electrodes. Primary electrons passing through the channels undergo multiplication, achieving a gain of several thousand. Owing to its finite bulk resistivity, the GMCP can effectively release accumulated positive charges, thereby suppressing charging effects and enhancing the operational stability of the GMPD.

-Topmetal-II chip: The detector's anode utilizes the Topmetal-II pixel chip, a silicon pixel detector produced using a 0.35 $\mu\mathrm{m}$ CMOS process. The chip consists of a 72×72 array of pixels within a central 6×6 mm² sensitive area, with a pixel spacing of 83 $\mu\mathrm{m}$. The total chip area is 8×9 mm². Each individual pixel has a top metal area of 15×15 $\mu\mathrm{m}$², which is exposed to directly collect charge. Each pixel electrode is connected to a charge-sensitive amplifier, which has an equivalent noise charge (ENC) of 13.9 e$^-$ at room temperature and even lower noise in a cryogenic environment, such as liquid nitrogen. For further details on the Topmetal-II chip, please refer to references \cite{an2016low, li2021preliminary}.

-Working gas: The working gas sealed inside the GMPD chamber is a mixture of 40\% helium (He) and 60\% dimethyl ether (DME) at 0.8 atm. DME is used to reduce diffusion during the electron drift process and also serves as a quenching agent. Compared with pure DME at the same pressure of 0.8 atm, the He+DME mixture produces slightly longer photoelectron tracks, while the addition of He also increases the GMCP gain, providing better signal amplification.

We have installed a BOSCH BME680 gas sensor inside the CXPD chamber to monitor the pressure and temperature of the working gas. These parameters are used to assess whether there is any significant gas leakage in the detector and can also be utilized for gain correction of the GMPD.

The GMPD utilizes a closed gas design, employing brazing and laser welding techniques to achieve a tight seal connection, ensuring high gas tightness and excellent mechanical performance. The detector features a 14 mm high drift region between the cathode and GMCP, and a 4 mm high induction region between the GMCP and the anode. The gas tightness of the GMPD is measured using a helium leak detector, with a gas leakage rate of less than $10^{-12}$\,Pa·m$^3$·s$^{-1}$.

To explore the technology of wide-FOV X-ray polarization observations, both GMPDs are designed with a large FOV. There are two coaxial through-holes with a diameter of 10 mm each on the scintillator detector, matching the size of the beryllium windows. The distance from the top surface of the CXPD CubeSat to the surface of the beryllium window is 37 mm. The FOV has a Full Width at Zero Response (FWZR) of 30.2°.

\begin{figure}
\centering
\includegraphics[scale=0.35]{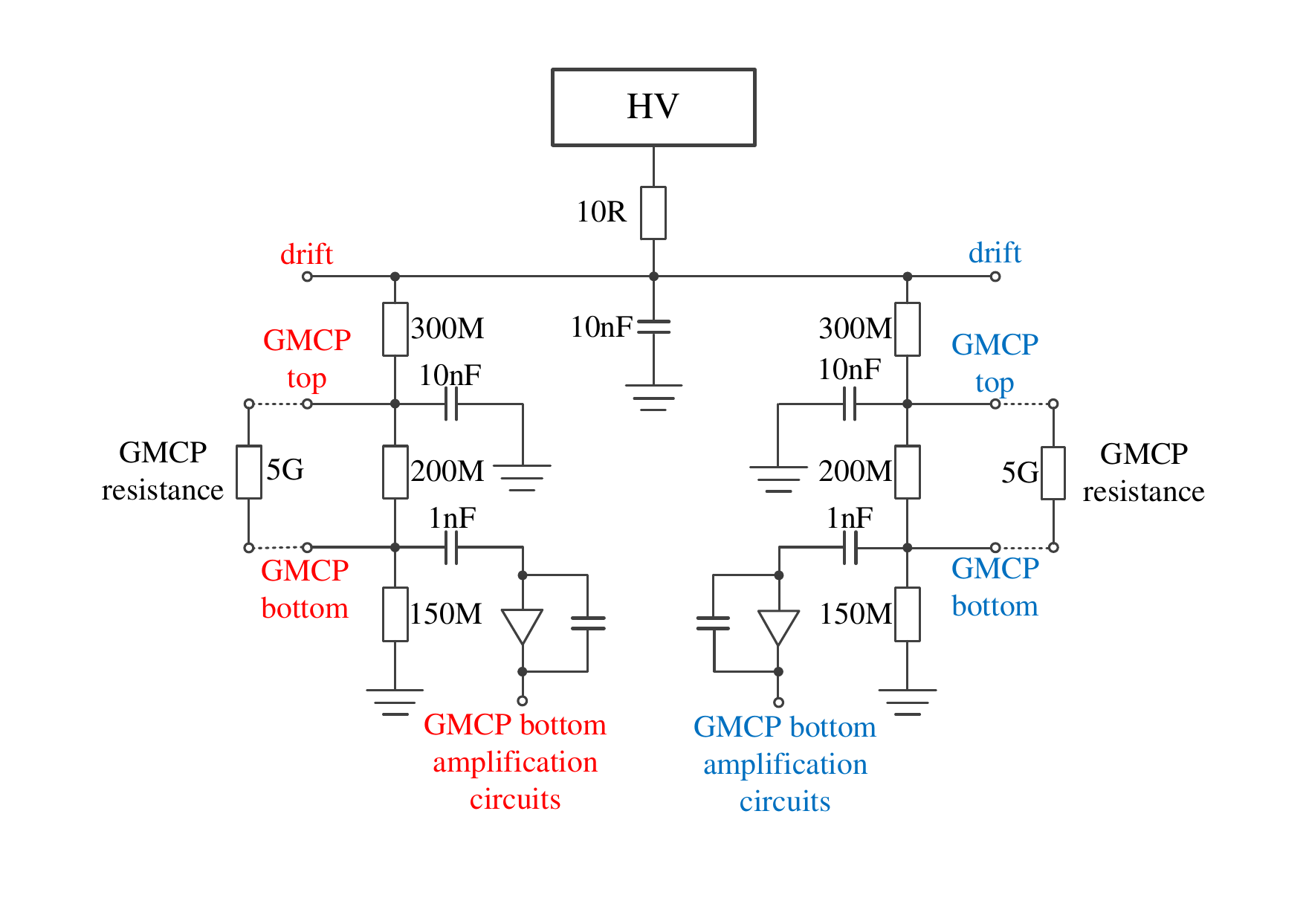}
\caption{High voltage circuits diagram.}
\label{fig:high_voltage}
\end{figure}

\subsection{High voltage}
The HV is supplied by the CAEM A7504 HV module, which is powered by a +12 V power source. The high voltage output can be programmed to vary from 0 to -4 kV and is monitored for actual voltage and current output through feedback pins. The high voltage output is divided into six independent channels, supplying power to the cathodes of the two GMPDs, the upper electrode, and the lower electrode of the GMCP. The voltage divider circuit is depicted in Fig.\ref{fig:high_voltage}. The HV module and the voltage divider circuit are arranged on the HV board of the electronic system. Electrical connections between the HV board and the HV electrodes of the GMCP are established using flying wires.

During the operation of the GMPD, a portion of the secondary electrons generated after the GMCP multiplication is collected by the bottom electrode of the GMCP. We have specially designed two bottom amplification circuits on the HV board to read out signals from the bottom electrode of the GMCP, thus enhancing the detector's time and energy resolution. For details on the bottom amplification circuit, please refer to the cited reference \cite{fan2023front}.

\begin{table*}
\centering
\caption{Types of data recorded by CXPD.}
\label{tab:payload}
\begin{tabular}{*{6}{l}}
\toprule
\multicolumn{1}{c}{Component} & \multicolumn{4}{c}{Data types}                                                         \\ \midrule
SiPM data                     & SiPM trigger time &                     &                        &                     \\
GMCP data                     & GMCP time         & GMCP ADC value      &                        &                     \\
Topmetal data                 & Topmetal time     & Pixel address value & Pixel ADC value        &                     \\
Monitoring data               & HV voltage value  & HV current value    & GMPD temperature value & GMPD pressure value \\ \bottomrule
\end{tabular}
\end{table*}

\subsection{Electronics system}
The FEE board primarily consists of the GMPD, SiPM trigger circuits, and GMPD data readout circuits. The BEE board includes the main controller, power regulator, data and firmware memory, communication interface and devices, and an external clock. The HV board comprises the GMCP bottom amplification circuit, dividing circuits, HV module, HV configuration, and HV monitoring. Inter-board communication and power are facilitated through Field-Programmable Gate Array (FPGA) mezzanine cards (FMC).

The signals from each detector are digitized, compressed, encoded, and stored before being transmitted to the ground. The electronic system employs a scheme where command and data transmission are separated to provide stable and high-speed communication while allowing for in-orbit upgrades of the electronic system. The electronic system can independently manage detector control and monitoring, high-speed data processing, and control of multiple devices.

The CXPD payload has only one external interface, which is sufficient to meet power supply, data transfer, and command transfer requirements. Reference \cite{wang2023electronics} provides detailed information on the specific functions and parameters of the electronic system.

\subsection{CXPD payload}
\begin{figure}
\centering
\includegraphics[scale=0.25]{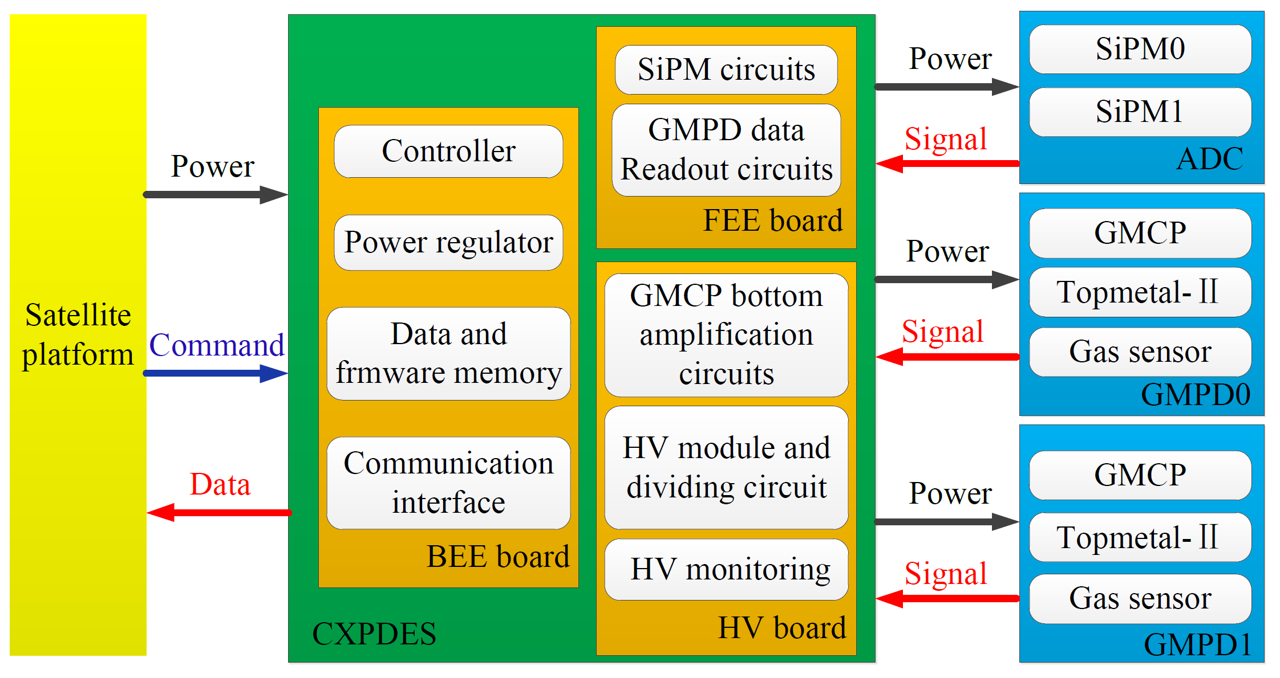}
\caption{The composition and functional module division of the CXPD payload.}
\label{fig:payload}
\end{figure}

The composition and functional module division of the CXPD payload are shown in Fig.\ref{fig:payload}. Firstly, the satellite platform provides 5 V and 12 V power supplies to the electronic system for main power and HV module dedicated power, respectively. The power module converts the 5 V power supply into various power sources distributed to different devices. Secondly, the electronic system receives commands from the satellite platform, configures the working parameters of SiPM trigger circuits, GMCP bottom amplification circuits, Topmetal chips, and HV module, and controls the devices to start (or stop) data acquisition, data storage, and transmission. Finally, the electronic system receives telemetry parameters such as voltage and current from the HV outputs, pressure and temperature from the gas sensor, and collects and processes measurement data from SiPM trigger circuits, GMCP bottom surface amplification circuits, and Topmetal chips.

Tab.\ref{tab:payload} summarizes the types of data recorded by the CXPD. For the Anti-Coincidence Detector, only the trigger time for each SiPM is recorded. For each GMCP bottom electrode signal, event arrival time and signal amplitude (GMCP ADC value) are recorded. For each Topmetal chip response, the response time, the address of the firing pixel, and the signal magnitude (pixel ADC value) are recorded. Engineering monitoring data, including HV voltage, HV current, GMPD temperature, and GMPD pressure values, are collected once per second. Reference \cite{wang2023electronics} provides detailed descriptions of the CXPD payload control logic, data processing, and data structure. 

The CXPD payload has dimensions of 10 × 10 × 12 cm$^3$, with a total mass of approximately 850 g. The typical power consumption during normal operation is around 4.5 W.

\section{Performance and Ground Calibration}\label{section3}
\subsection{Detection efficiency}

\begin{figure}
\centering
\includegraphics[scale=0.42]{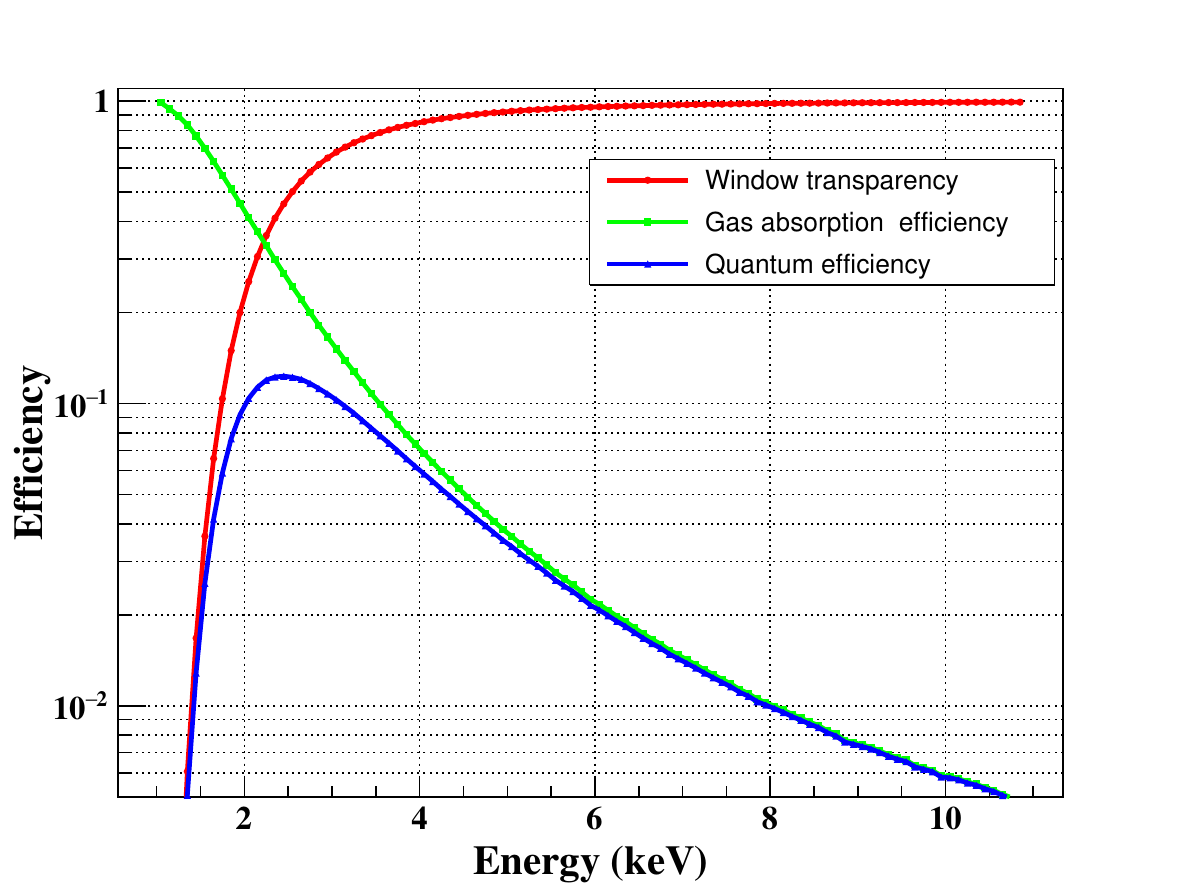}
\caption{GMPD quantum efficiency as a function of energy.}
\label{fig:detection_efficiency}
\end{figure}

The detection efficiency of the GMPD is determined by the beryllium window (100 $\mu\mathrm{m}$) and the working gas (a 1.4 cm thick mixture of 40\% He and 60\% DME, at a pressure of 0.8 atm). We used a simulator \cite{huang2021simulation} to calculate the detection efficiency, as shown in Fig.\ref{fig:detection_efficiency}. This geometry provides a peak detection efficiency of over 12\% at 2.5 keV, which decreases to approximately 0.6\% at 10 keV. In addition, simulation results indicate that the number of secondary photons produced by the anti-coincidence detector and entering the main detector is negligible, and therefore they are not included in the detector efficiency curve.

\begin{table}
\centering
\caption{The measured energy resolutions.}
\label{tab:spec}
\begin{tabular}{*{6}{l}}
\toprule
\multicolumn{1}{c}{\multirow{2}{*}{E (keV)}} & \multicolumn{4}{c}{FWHM/E}                                                \\ \cmidrule(l){2-5} 
\multicolumn{1}{c}{}                         & \multicolumn{2}{c}{GMPD0}           & \multicolumn{2}{c}{GMPD1}           \\ \midrule
2.98                                         & \multicolumn{2}{l}{0.3063 ± 0.0007} & \multicolumn{2}{l}{0.3124 ± 0.0007} \\
4.51                                         & \multicolumn{2}{l}{0.2458 ± 0.0006} & \multicolumn{2}{l}{0.2542 ± 0.0006} \\
6.40                                         & \multicolumn{2}{l}{0.2086 ± 0.0005} & \multicolumn{2}{l}{0.2145 ± 0.0005} \\
8.05                                         & \multicolumn{2}{l}{0.1775 ± 0.0004} & \multicolumn{2}{l}{0.1801 ± 0.0005} \\ \bottomrule
\end{tabular}
\end{table}

\subsection{Energy calibration}
\begin{figure*}
\centering
\includegraphics[scale=0.52]{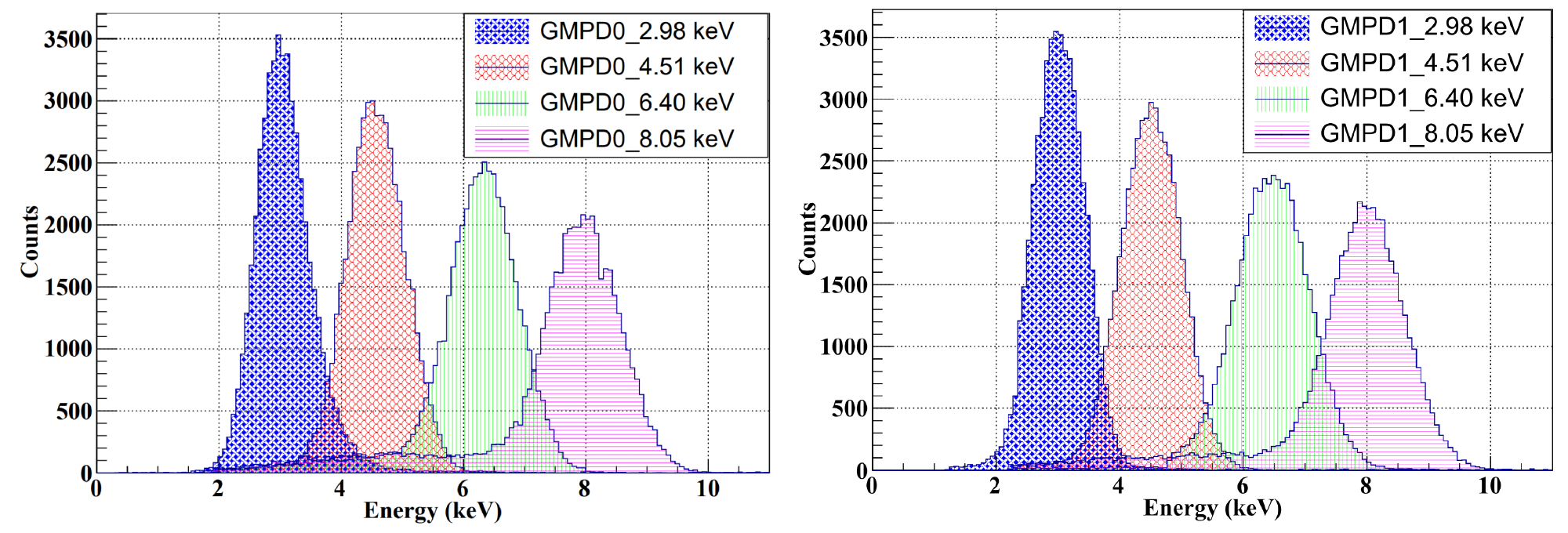}
\caption{Energy spectra measured with 45-degree Bragg diffraction. Left: GMPD0. Right: GMPD1.}
\label{fig:energy_resolution}
\end{figure*}

We used four Bragg crystals and their 45-degree diffraction to measure the energy spectra and modulation factors of the two GMPDs \cite{xie2023variably}. A silicon drift detector was used to detect the diffracted spectra and verify if the peaks appeared at the expected energies. Fig.\ref{fig:energy_resolution} shows the pulse height spectra of the diffracted beam. Some photons are absorbed by the entrance window or the Ni-Cr alloy on the surface of the GMCP, generating photoelectrons that exit the material and deposit some of their energy in the working gas. These events are characterized by incomplete charge collection, which dominates the low-energy tail in the pulse height spectrum. To extract spectral information, Gaussian functions were fitted to each peak, and the results are shown in Tab.\ref{tab:spec}.

\begin{figure}
\centering
\includegraphics[scale=0.29]{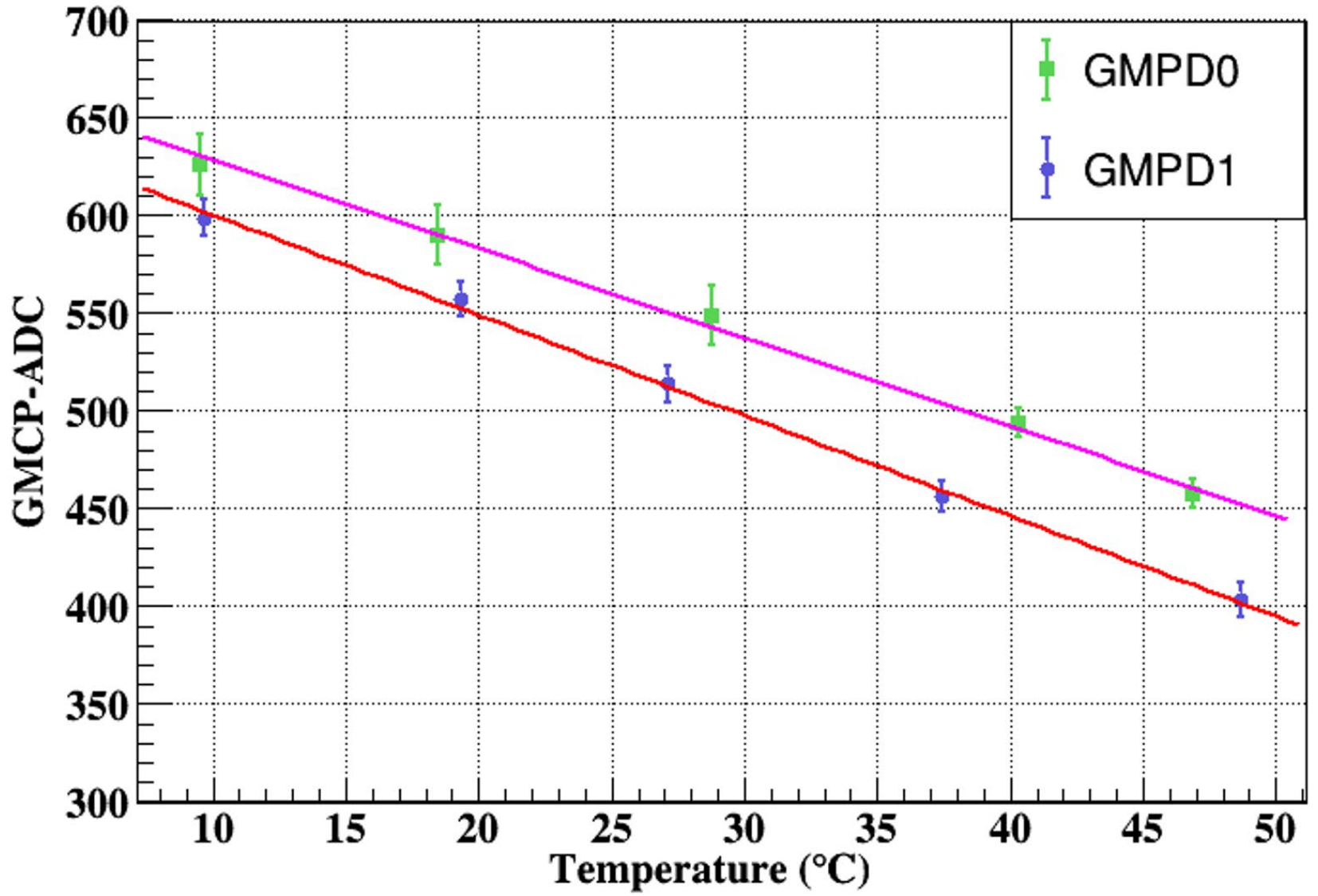}
\caption{The pulse height of the photopeak versus the environmental temperature.}
\label{fig:adc_tempature}
\end{figure}

Although GMCP with bulk resistance can eliminate the charging-up effect of the gas detector and provide operational stability, the bulk resistance of GMCP decreases with increasing environmental temperature. Since the voltage divider circuit is used to supply high voltage to different electrodes of the GMPD, changes in GMCP bulk resistance can cause variations in the voltage at both ends of the GMCP, ultimately resulting in changes in GMPD gain. Therefore, calibration experiments were conducted on both GMPD detectors at different environmental temperatures on the ground. Using $^{55}$Fe as the X-ray source and an operating voltage of 3200 V, the measurement results, as shown in Fig.\ref{fig:adc_tempature}, indicate that the gain decreases with increasing temperature. We performed a linear fit and found that the pulse height (gain) of the energy spectrum is linearly related to temperature.

\begin{figure}
\centering
\includegraphics[scale=0.29]{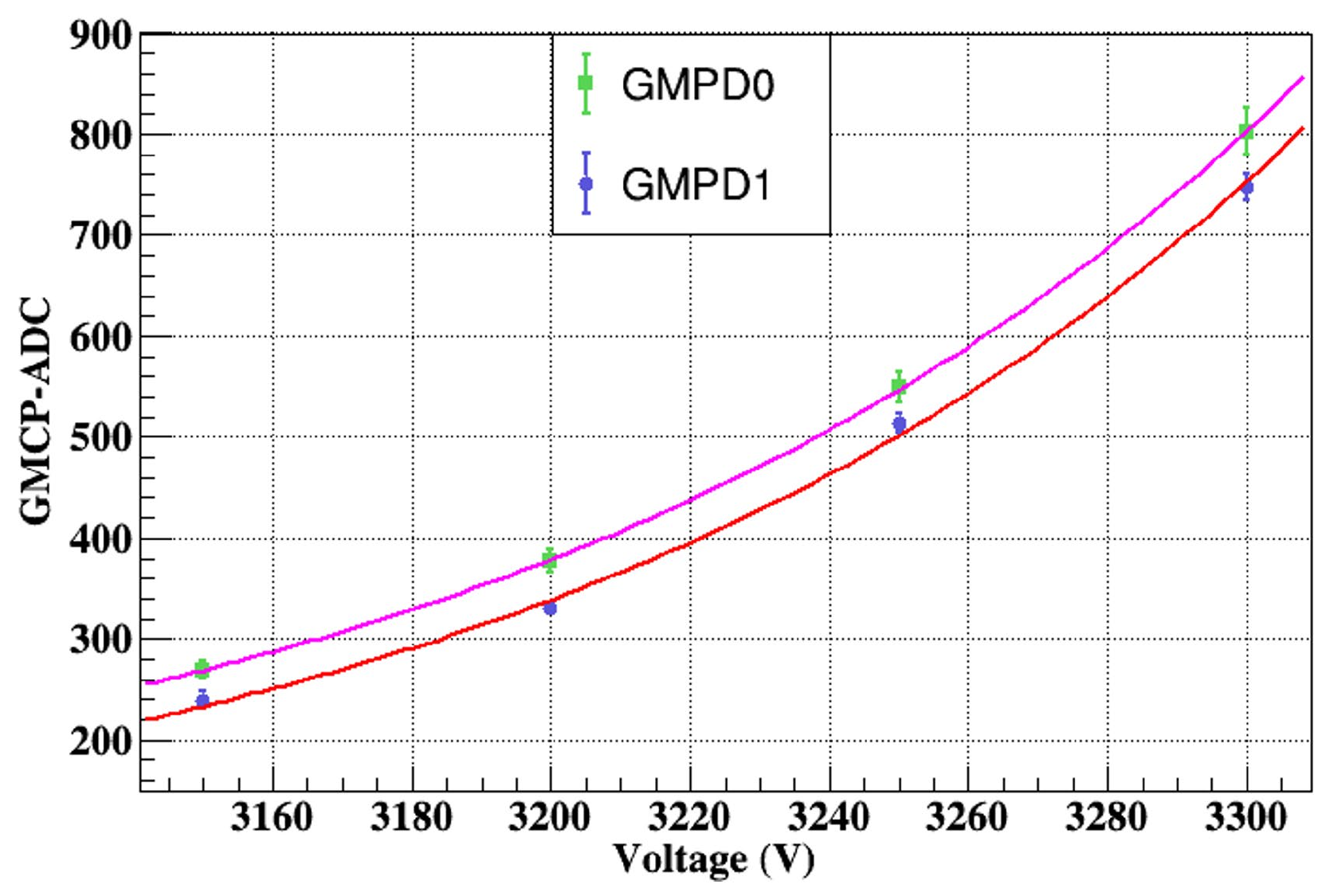}
\caption{The pulse height of the photopeak versus the voltage value.}
\label{fig:adc_v}
\end{figure}

Additionally, changes in the voltage values of the high-voltage module were made to measure the variation in gain, and the results are shown in Fig.\ref{fig:adc_v}. We used an exponential function for fitting, and with increasing voltage, the GMPD gain exhibits exponential growth.

\subsection{Polarization detecting performance}

\begin{table*}
\centering
\caption{The values of modulation degree at different energy, and the systematic error $\sigma_{\text{sys}}$, statistical error $\sigma_{\text{stat}}$.}
\label{tab:compare}
\begin{tabular}{*{6}{c}}
\toprule
Energy  & Polarization degree & Modulation/Residual &$\sigma_{\text{sys}}$ &$\sigma_{\text{stat}}$ & $\sigma_{\text{total}}$  \\ \hline
\multirow{2}{*}{2.98\,keV}  & 97.4\% & 0.285 & \multirow{2}{*}{0.001}              & 0.008     & 0.008                     \\
                         &   0.0 & 0.008                     &                    & 0.007     & 0.007                     \\
\multirow{2}{*}{4.51\,keV}  & 99.8\% & 0.474 & \multirow{2}{*}{0.001}              & 0.003     & 0.003                     \\
                         &   0.0 & 0.009                     &                    & 0.006     & 0.006                    \\
\multirow{2}{*}{6.40\,keV}  & 99.8\% & 0.606 & \multirow{2}{*}{0.002}              & 0.004     & 0.004                      \\
                         &   0.0 & 0.008                     &                    & 0.006     & 0.006                     \\
\multirow{2}{*}{8.05keV}    & 99.8\% & 0.612 & \multirow{2}{*}{0.002}              & 0.003     & 0.003                     \\
                         &   0.0 & 0.004                     &                    & 0.004     & 0.005                     \\ \hline
\end{tabular}
\end{table*}

Since 45-degree diffraction generates completely polarized X-ray beams, the same data were used to calculate the modulation factors. The emission angle distribution of electrons was fitted with the modulation function
\begin{equation}
\label{modu_equa}
\mathcal{M}(\varphi) = A + B \cos^2(\varphi - \varphi_0)
\end{equation}
where $A$ and $B$ are constants. The phase of the cosine, $\varphi_0$, singles out the direction where the emission is more probable and therefore represents the polarization angle.

Subsequently, we can obtain the modulation factor $\mu$ 
\begin{equation}
\mathcal{\mu}=\frac{\mathcal{M}_{\max }-\mathcal{M}_{\min }}{\mathcal{M}_{\max }+\mathcal{M}_{\min }}=\frac{B}{2 A+B}
\end{equation}

where $\mathcal{M}_{\max }$ and $\mathcal{M}_{\min }$ are the maximum and minimum values of the modulation function, respectively. 

 Both detectors, after calibration and correction, exhibit residual modulation levels below 1\% at several energy points within the 3–8 keV range \cite{yi2024effectiveness}, with system errors controlled to below 0.3\%. In terms of its capability to detect polarized sources, the CXPD achieves a maximum modulation factor of 61.2\%, which is higher than that of IXPE in the energy range above 5 keV \cite{di2022calibration}, as shown in Tab.~\ref{tab:compare} using CXPD01 as an example. However, due to the use of a He–DME gas mixture, its detection efficiency is lower than that of IXPE. Additionally, the CXPD maintains consistent modulation reconstruction for sources at different polarized phases, and demonstrates a strong linear relationship between polarization degree and modulation factor.

\section{In-orbit status and observations}\label{section4}
\subsection{In-orbit operation}
\begin{figure}
\centering
\includegraphics[scale=0.4]{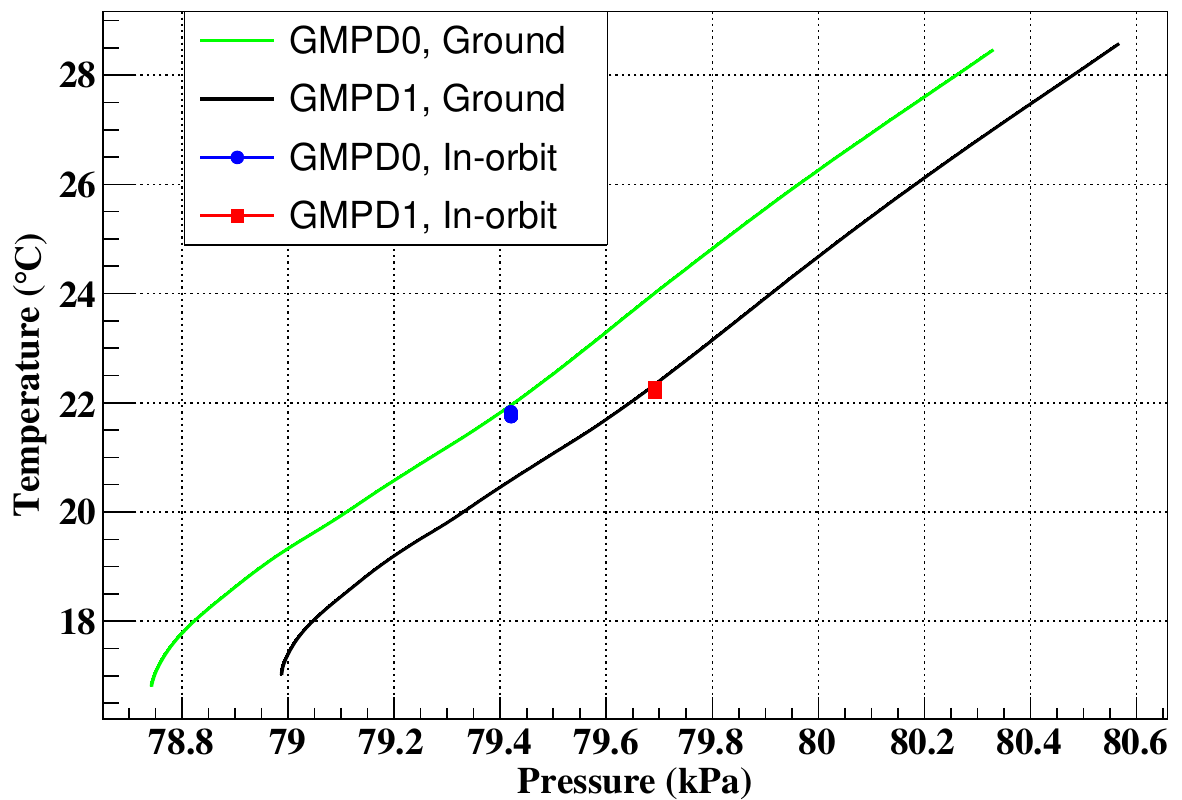}
\caption{State of the working gas in the two gas detectors.
}
\label{fig:in_orbit}
\end{figure}

On June 11, 2023, the CXPD payload underwent its first brief power-on test, primarily aimed at checking the status of the gas detectors, including temperature and pressure within the sealed chambers. The results, as shown in Fig.\ref{fig:in_orbit}, successfully monitored the status parameters of both GMPD detectors. These data matched the ground calibration data, confirming that there were no leaks in the working gas of the two gas detectors and validating the reliability of the detector's encapsulation technology.

Subsequently, on August 8th, the high voltage was applied to the set value, and GMPD successfully collected the tracks generated by X-ray photoelectrons and charged particles. This indicates that CXPD is operating normally in orbit and has completed the technical verification of X-ray polarization measurements. 

During its orbit, the CXPD conducted polarization observations of two sources: the first was Sco X-1, the first discovered extrasolar X-ray source and the brightest persistent object in the keV sky besides the Sun, and the second was Swift J1727.8-1613, a low-mass X-ray binary black hole that underwent a bright outburst detected on August 24, 2023. The CubeSat utilized a star tracker for attitude determination, achieving a pointing accuracy of approximately 0.02°, which ensured stable alignment with the target sources during observations.

\subsection{Observations and Data Reduction}
\label{sec: Observations and Data Reduction}
\begin{figure*}
\centering
\includegraphics[scale=0.24]{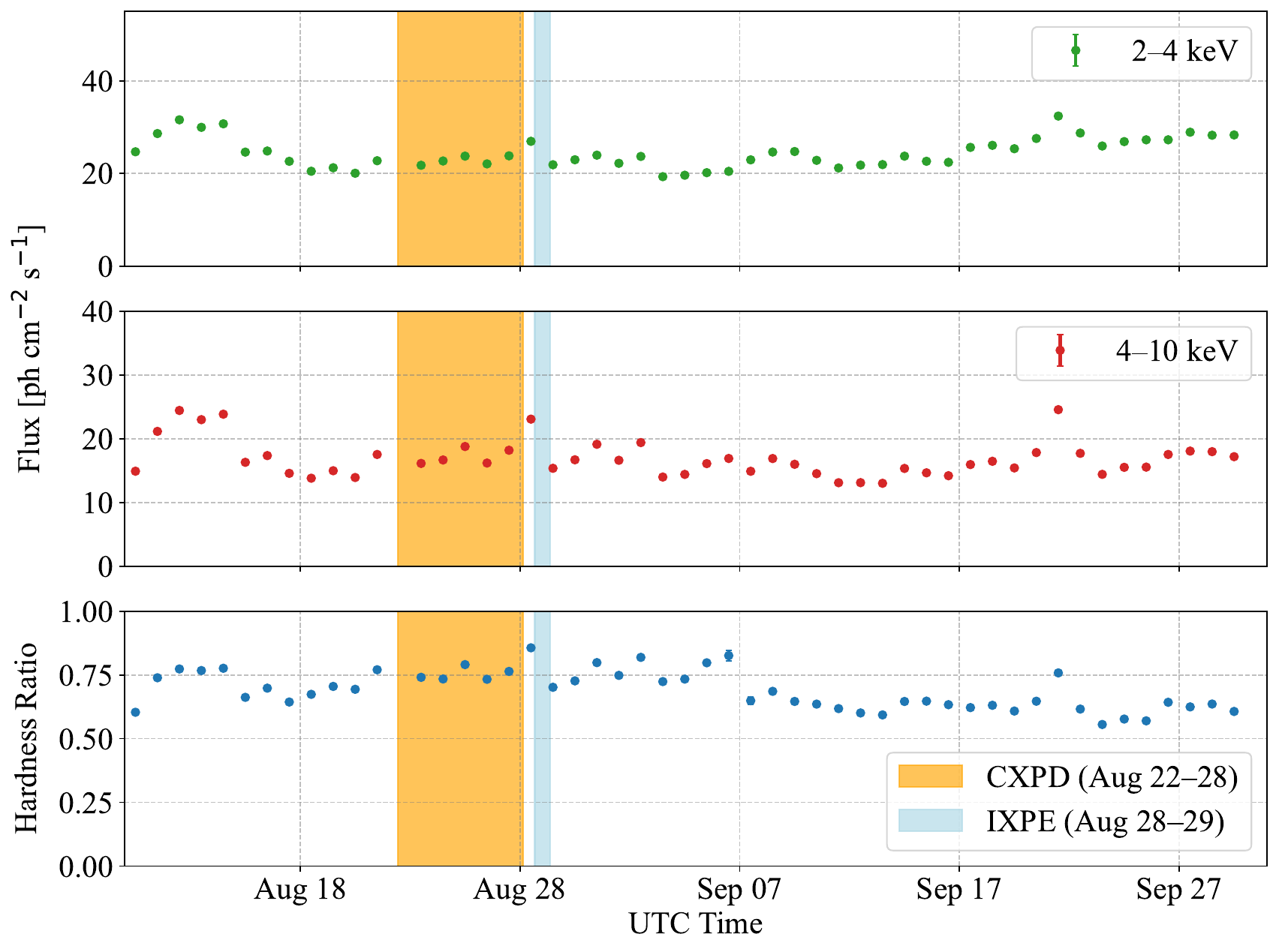}
\includegraphics[scale=0.2]{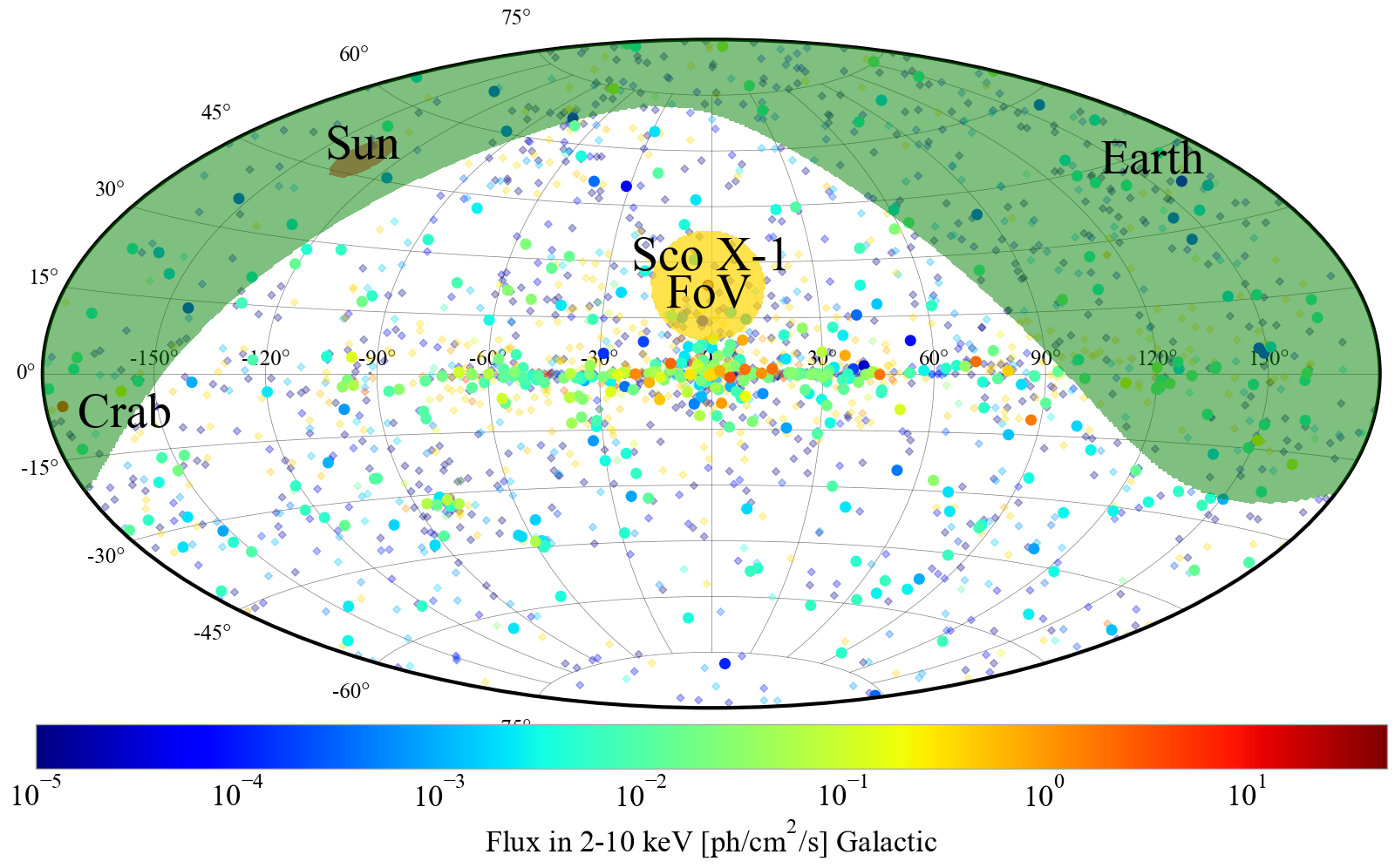}
\caption{Left: Observation period of the CXPD (orange shading) and the source flux of Sco X-1 measured by MAXI from 2023-08-22 10:24:14 UTC to 2023-08-28 03:39:04 UTC, with the blue shading indicating the most recent IXPE observation intervals. Right: Detector's pointing direction in Galactic coordinates at a specific time during the observation period, with yellow shading representing the FOV, while red and green shadings indicate Earth and the Sun, respectively.}
\label{fig:sco_time}
\end{figure*}

From 2023-08-22 10:24:14 UTC to 2023-08-28 03:39:04 UTC, the detector was pointed at Sco X-1 for observation. The left panel of Fig.\ref{fig:sco_time} shows the observation period of the CXPD and the source flux measured by MAXI \cite{matsuoka2009maxi} during this time, with the blue shading indicating the most recent IXPE observation intervals. The total effective exposure on Sco~X-1 during this period is 19,806~s. The right panel of Fig.\ref{fig:sco_time} illustrates the detector's pointing in Galactic coordinates. In the figure, the yellow shading represents the detector's FOV, while the red and green shading represent Earth and the Sun, respectively.

\begin{figure*}
\centering
\includegraphics[scale=0.24]{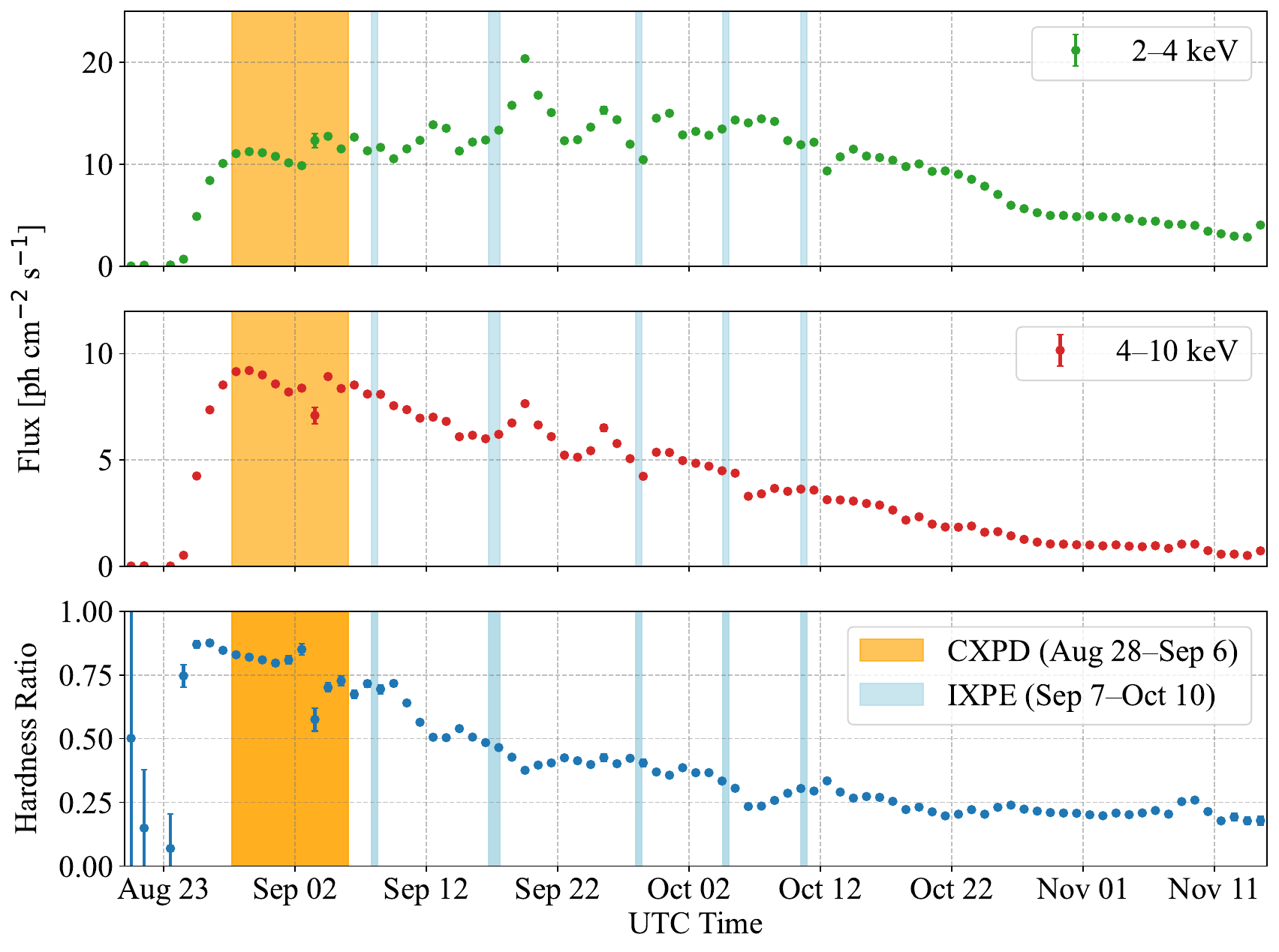}
\includegraphics[scale=0.2]{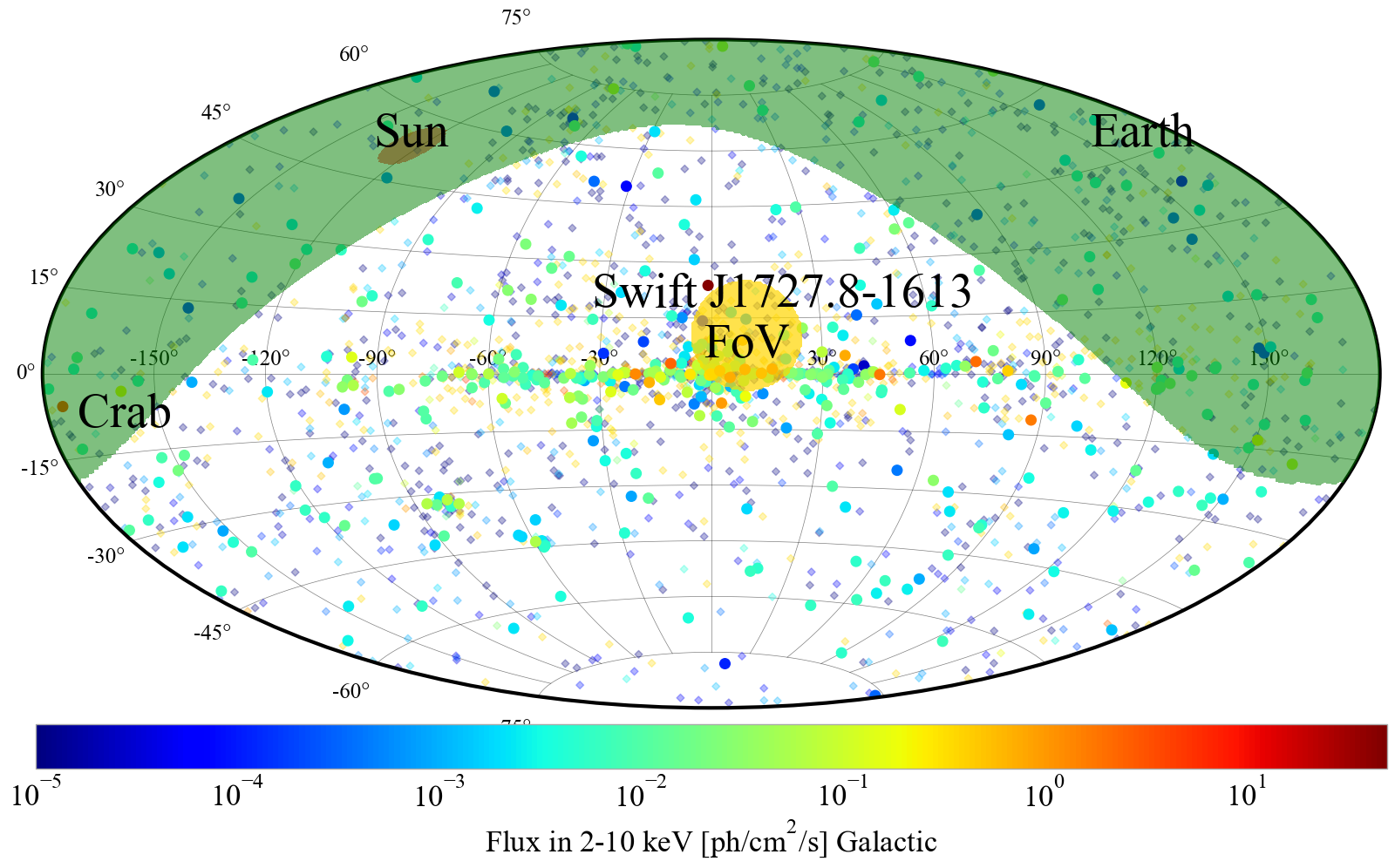}
\caption{Left: Observation period of Swift J1727.8-1613 from 2023-08-28 03:45:45 UTC to 2023-09-06 00:38:45 UTC, with the blue shading highlighting the simultaneous observation period of IXPE during the source's outburst. Right: Detector's pointing direction in Galactic coordinates at a specific time during the observation period as Fig.\ref{fig:sco_time}.}
\label{fig:swift_time}
\end{figure*}

Similarly, the observation time and pointing of Swift J1727.8-1613 are shown in Fig.\ref{fig:swift_time}, with the observation period from 2023-08-28 03:45:45 UTC to 2023-09-06 00:38:45 UTC. Additionally, IXPE performed five subsequent observations of the outburst, with all corresponding intervals indicated by the blue shading in the figure. The total effective exposure on Swift~J1727.8$-$1613 amounts to 49,814~s.

The data transmitted from the CubeSat includes the detector's position coordinates in space, temperature and pressure monitoring data, ADC values and timestamp information for threshold-crossing events on the GMCP lower surface, the topmetal track images, and the corresponding timestamp information for each image. Before conducting spectral and polarization analysis, the data must first undergo preprocessing, which includes event matching for Topmetal and GMCP, removal of edge events from the chip, clustering algorithms, selection of the satellite's position in the Earth's projection, and track feature extraction.

For the CXPD's 30.2° FOV, when conducting polarization observations of the two sources mentioned above, the primary background components in space are cosmic diffuse X-rays, high-energy charged particles \cite{xie2021study, huang2021modeling}, and other bright X-ray sources entering the FOV \cite{feng2023orbit}. High-energy charged particles can pass directly through the detector, so their count rate is not significantly affected by the FOV size. Furthermore, we can remove charged particles through energy and track morphology analysis \cite{xie2021study, feng2023orbit, zhu2021discrimination}. 
\begin{figure*}
\centering
\includegraphics[scale=0.4]{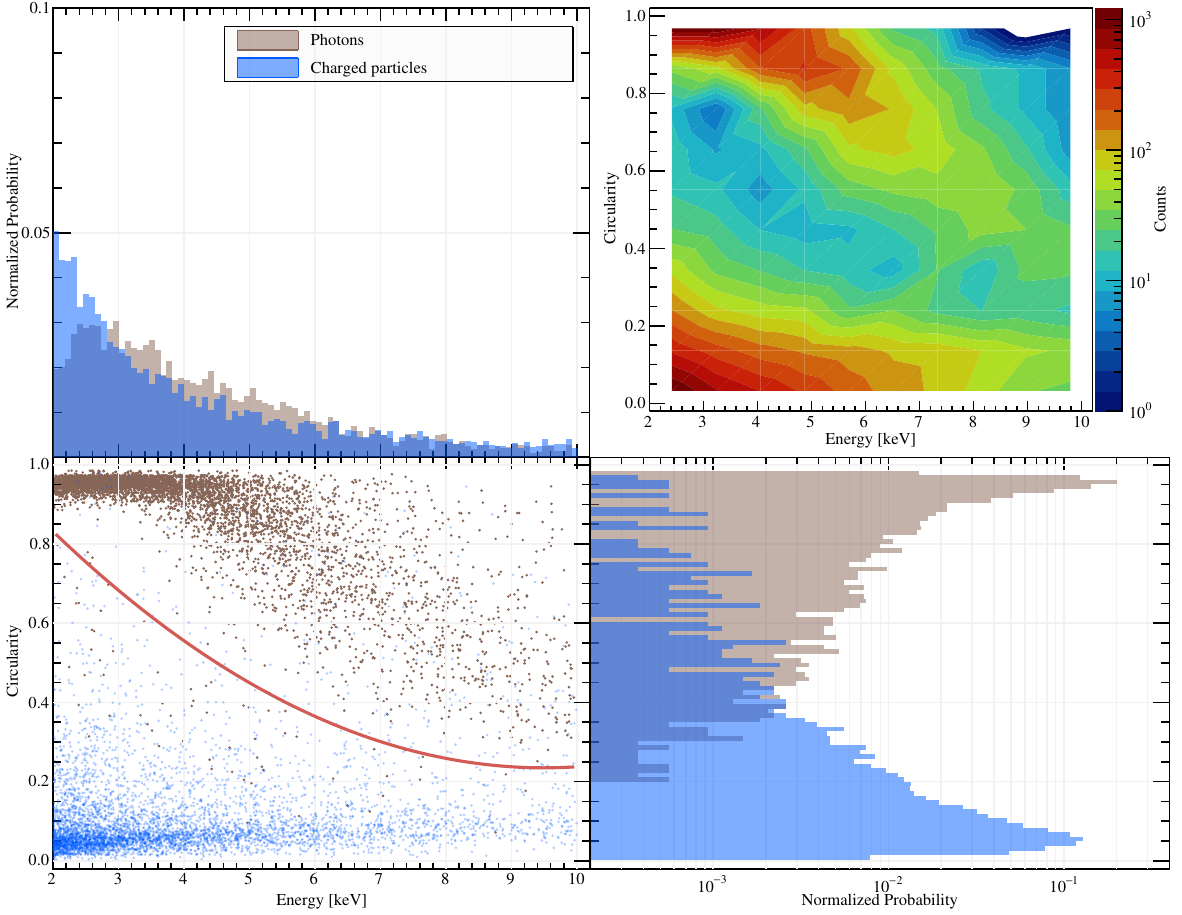}
\includegraphics[scale=0.16]{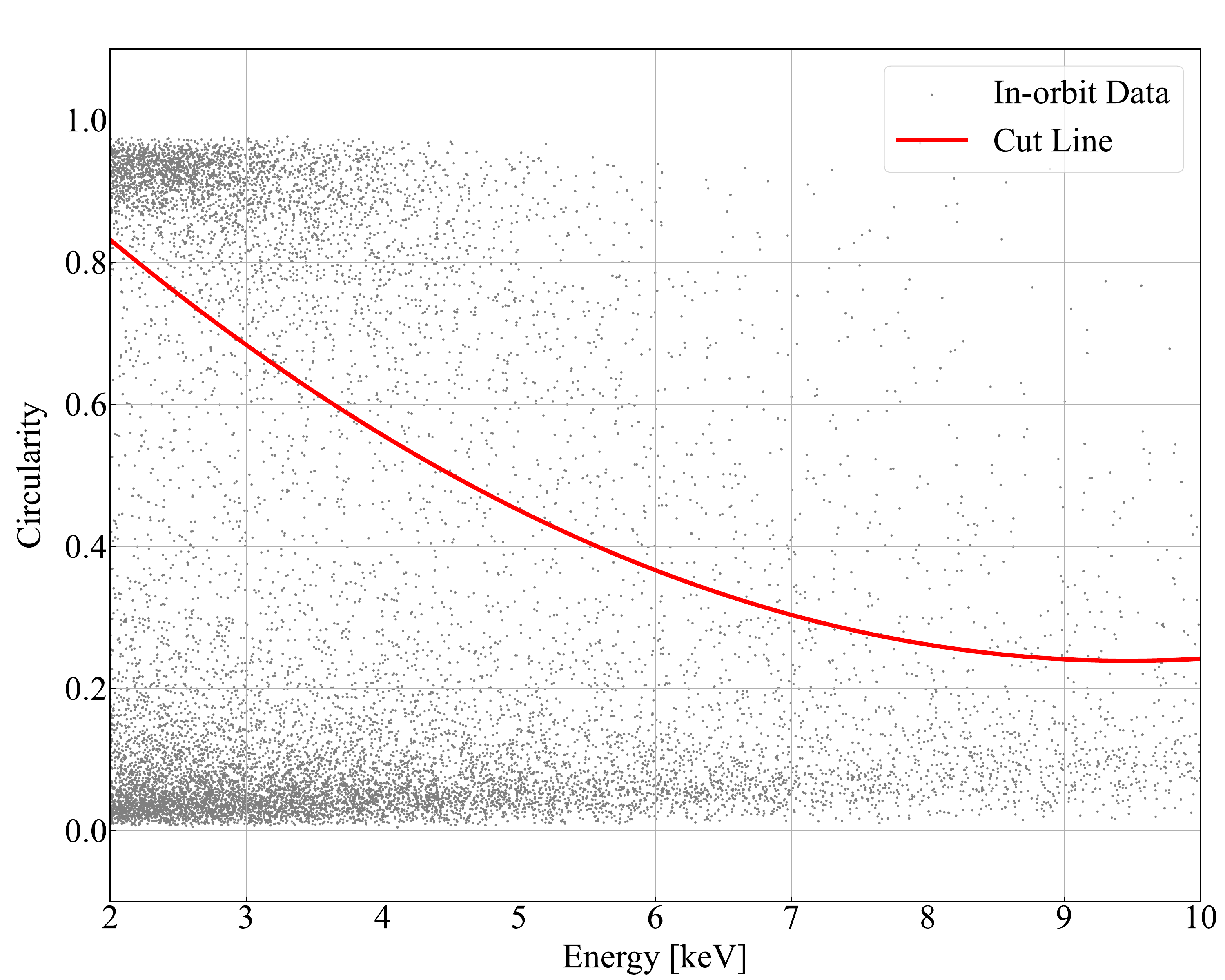}
\caption{The algorithm of charged-particle background screening. Left: Simulation results. The equation of the red line is \( y = 0.0106x^2 - 0.2014x + 1.1916 \). The background rejection rate after the linear cut is 94.06\%, and the photon retention rate is 98.81\%. Right: In-orbit measurement results, where the red cut line is the same as in the left panel.}

\label{fig:adc_cir}
\end{figure*}

\begin{figure*}
\centering
% \includegraphics[scale=0.095]{Pic/sco_pho.pdf}
% \put(-340, 20){\fontsize{10}{14}\selectfont \textbf{(a)}} 
% \includegraphics[scale=0.095]{Pic/sco_ele.pdf}
% \includegraphics[scale=0.095]{Pic/swift_pho.pdf}
% \includegraphics[scale=0.095]{Pic/swift_ele.pdf}

% \put(-340, 240){\fontsize{10}{14}\selectfont \textbf{(b)}} 
% \put(-528, 120){\fontsize{10}{14}\selectfont \textbf{(c)}}
% \put(-340, 120){\fontsize{10}{14}\selectfont \textbf{(d)}} 
    % \centering
\begin{minipage}{0.48\textwidth}
    \includegraphics[scale=0.095]{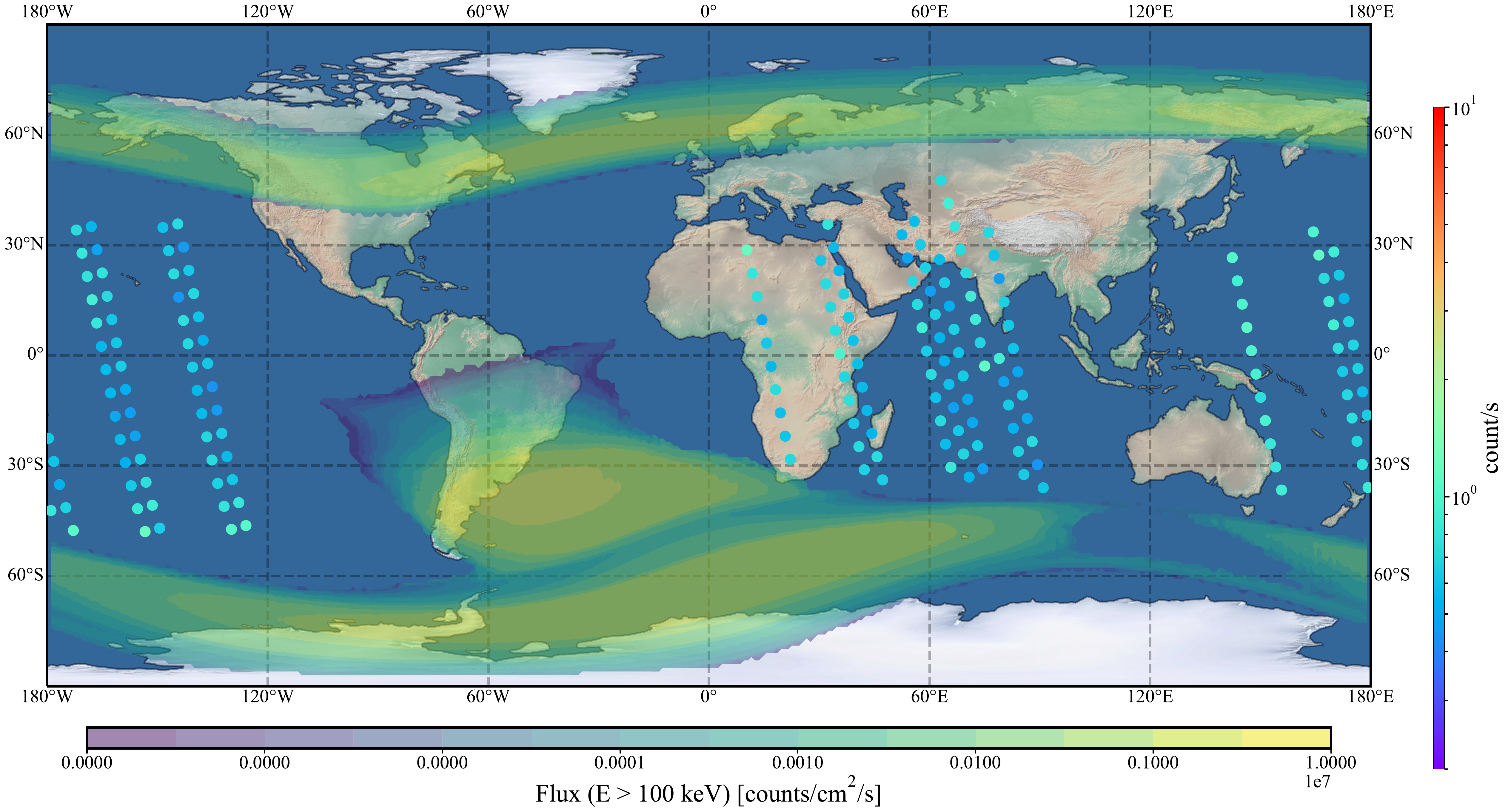}
    \put(-243, 120){\fontsize{10}{14}\selectfont \textbf{(a)}}
\end{minipage}
\begin{minipage}{0.48\textwidth}
    \includegraphics[scale=0.095]{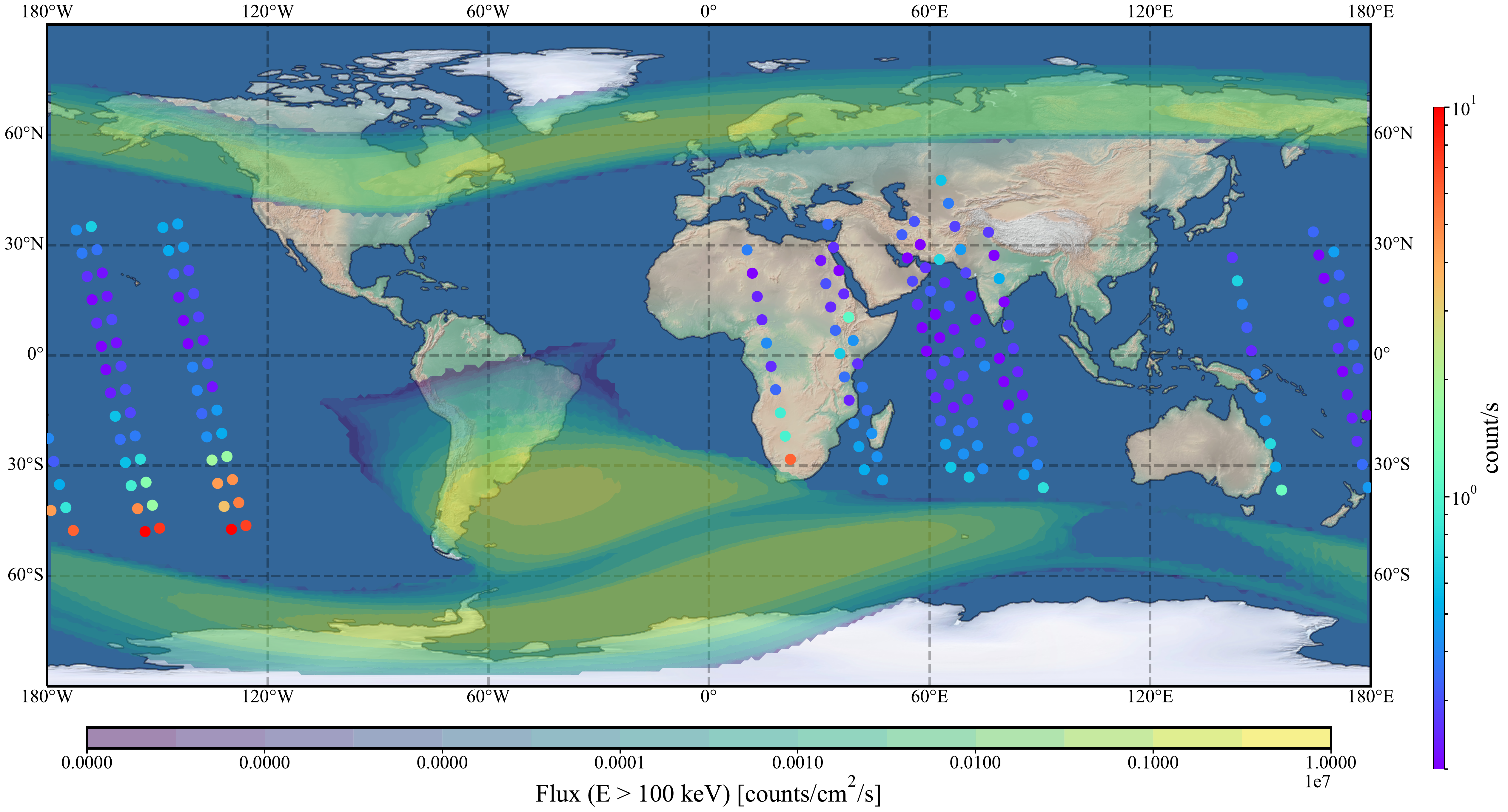}
    \put(-243, 120){\fontsize{10}{14}\selectfont \textbf{(b)}}
\end{minipage}

\vspace{5pt} % 调整上下图间距

\begin{minipage}{0.48\textwidth}
    \includegraphics[scale=0.095]{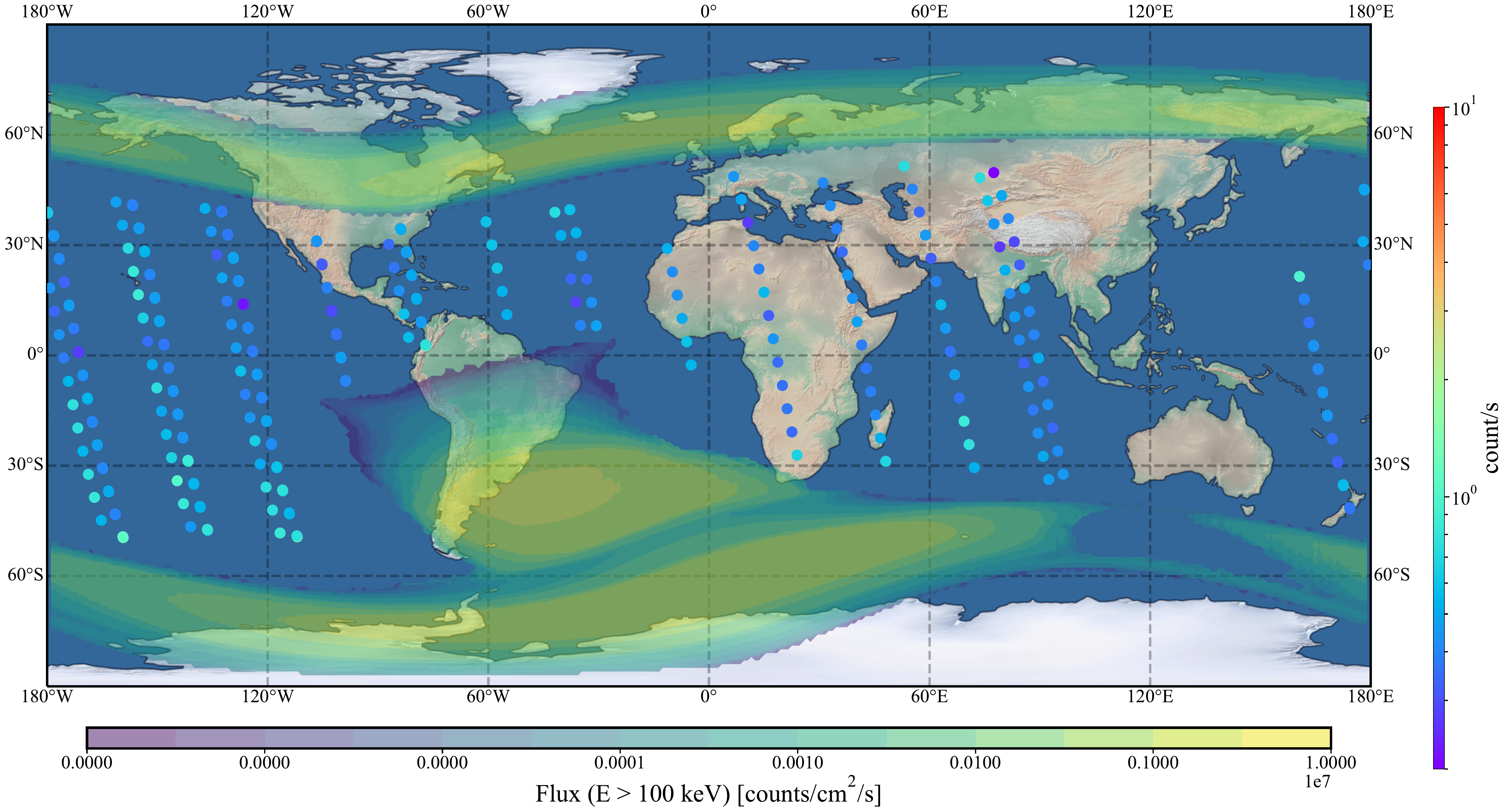}
    \put(-243, 120){\fontsize{10}{14}\selectfont \textbf{(c)}}
\end{minipage}
\begin{minipage}{0.48\textwidth}
    \includegraphics[scale=0.095]{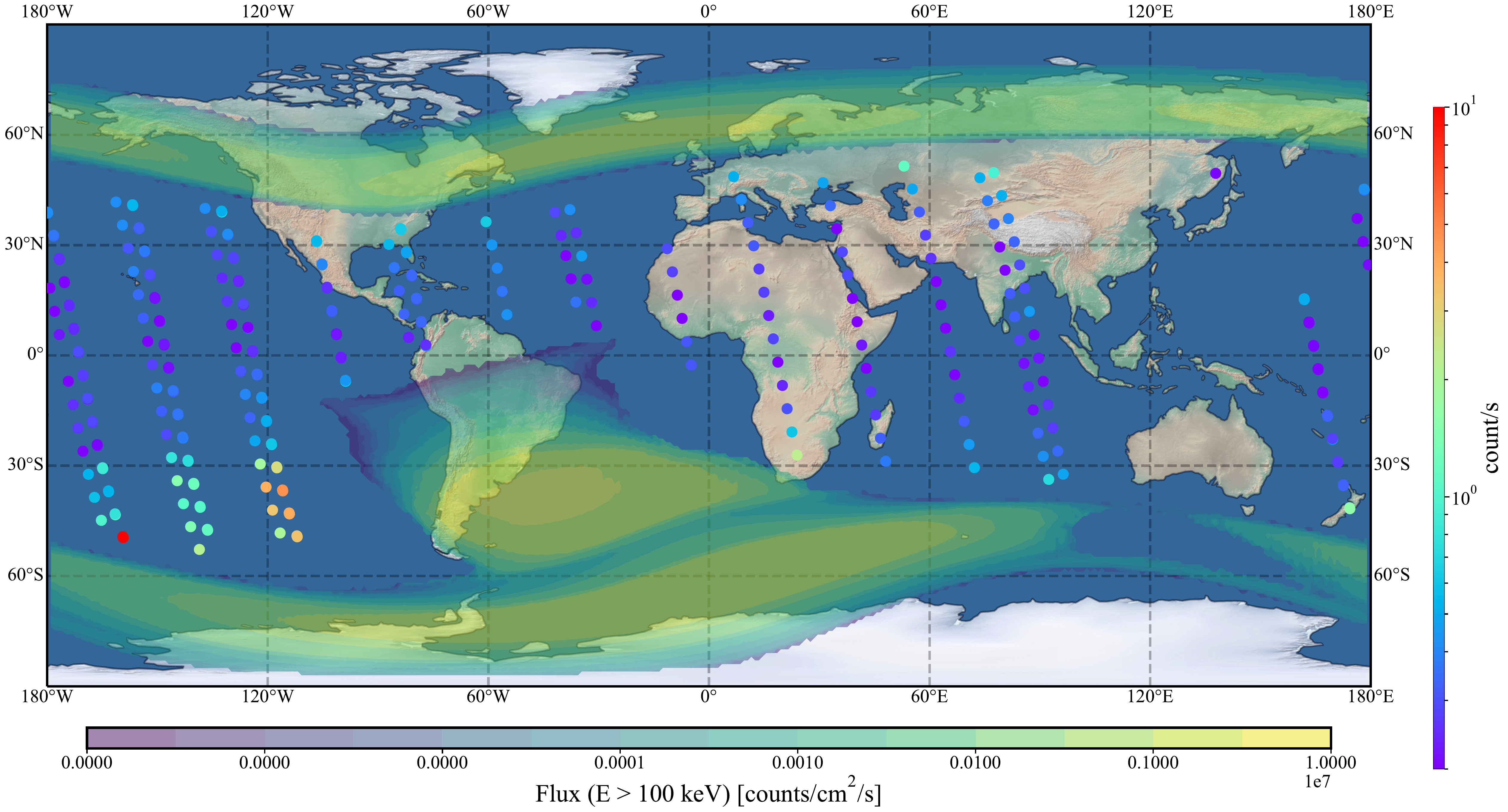}
    \put(-243, 120){\fontsize{10}{14}\selectfont \textbf{(d)}}
\end{minipage}
\caption{Photoelectron and charged particle count rates after the selection algorithm (as shown in Figure \ref{fig:adc_cir}), varying with position projection on earth's surface of the cubesat: (a) photon count rate during observations of Sco X-1, (b) charged particle count rate during observations of Sco X-1, (c) photon count rate during observations of Swift J1727.8-1613, (d) charged particle count rate during observations of Swift J1727.8-1613. The colormap shows the flux of trapped electrons with energies above 100 keV in the orbit.}
\label{fig:distinguishing}
\end{figure*}

\begin{figure*}
\centering
\includegraphics[scale=0.4]{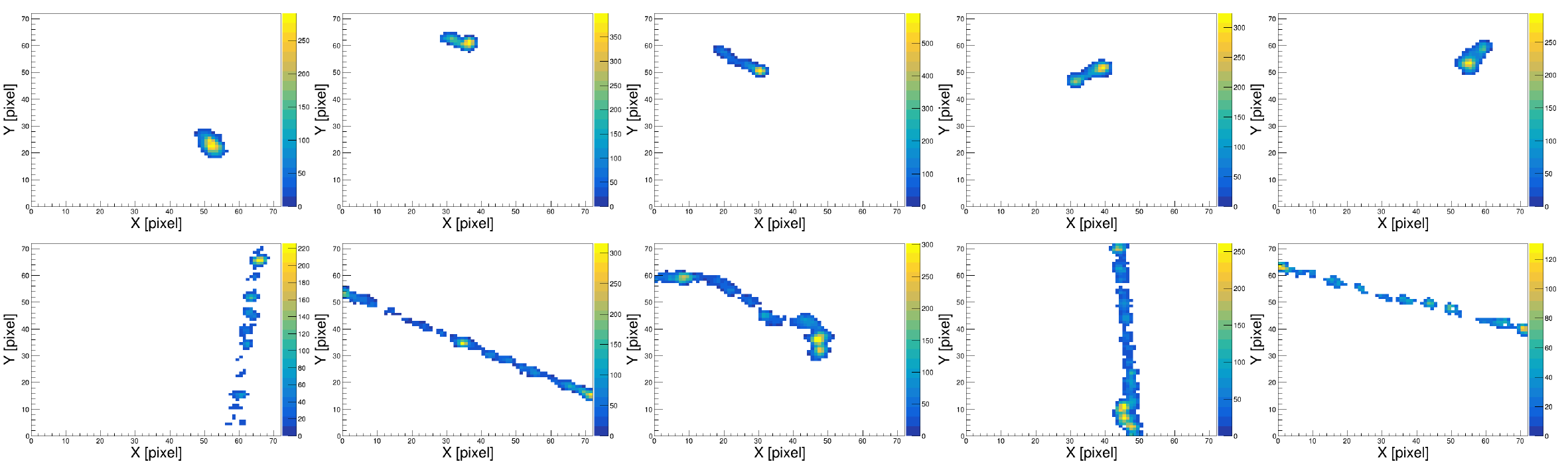}
\caption{In-orbit measurements of photoelectron tracks. The top row illustrates photon tracks identified by the track screening algorithm, while the bottom row displays high-energy charged particle tracks.}
\label{fig:track_image}
\end{figure*}

\begin{figure*}
\centering
\includegraphics[scale=0.4]{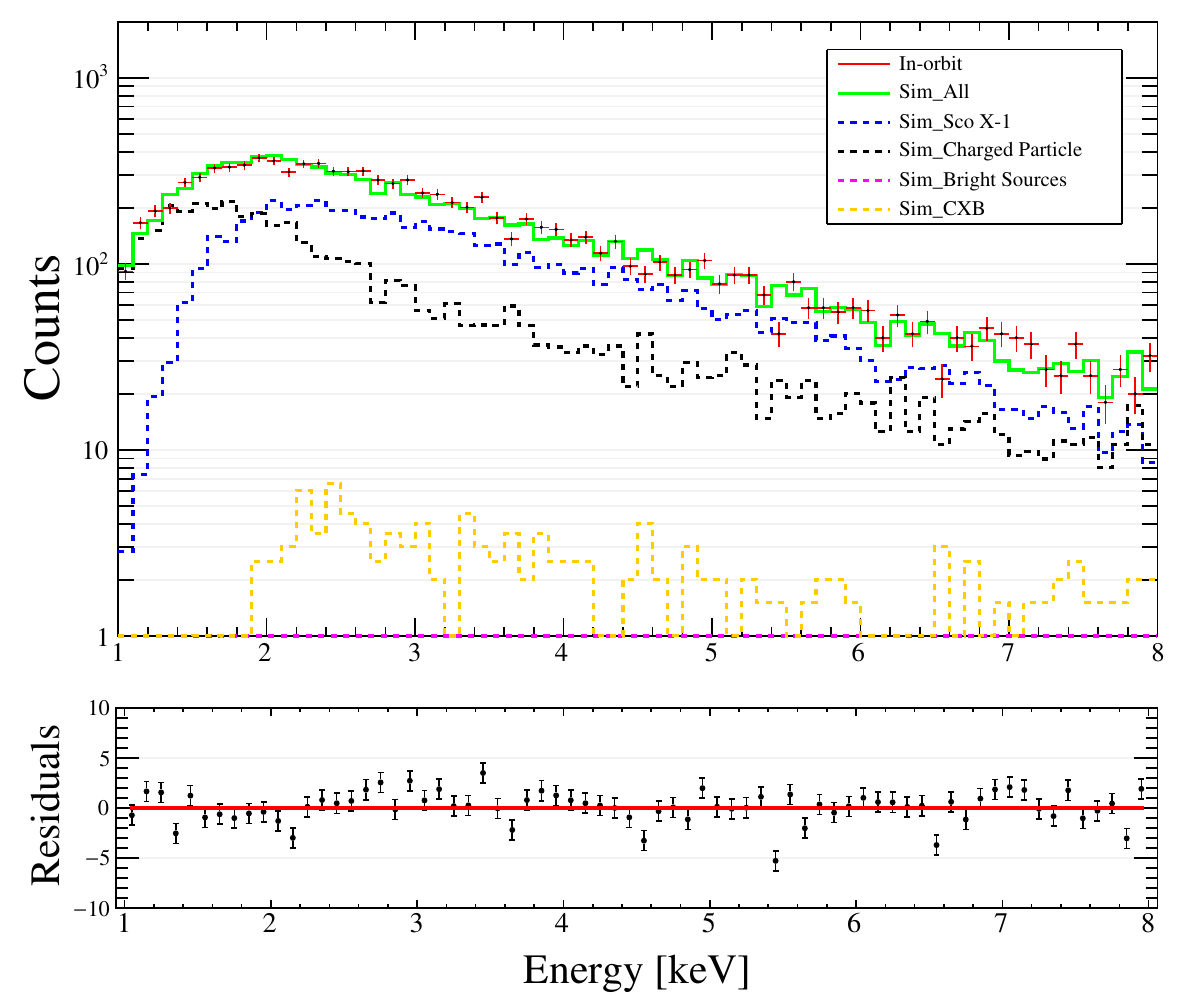}
\includegraphics[scale=0.4]{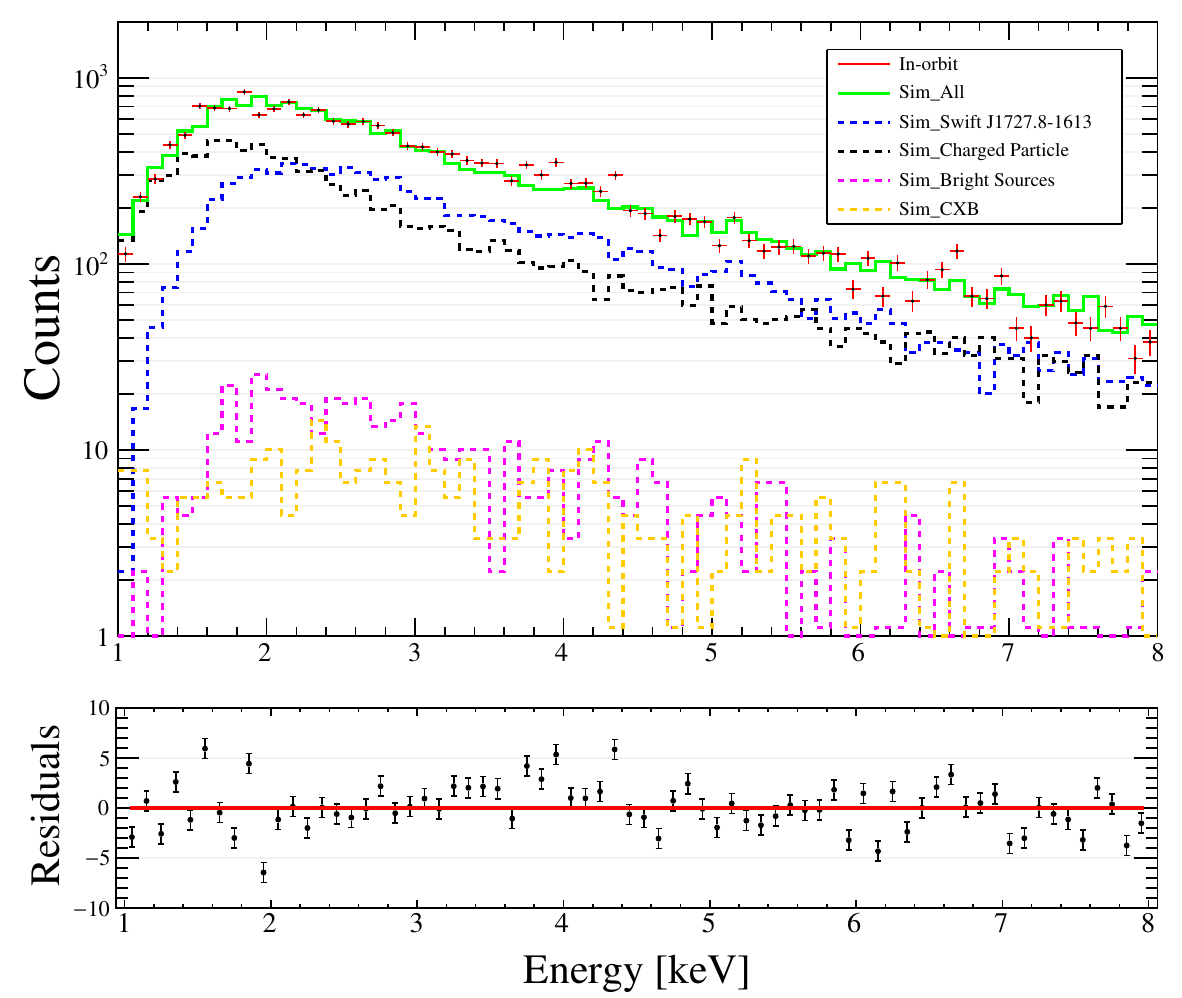}

\caption{Measured and simulated spectrum during the observation of Sco X-1 and Swift J1727.8-1613. Left: Sco X-1, Right: Swift J1727.8-1613. The red points with errors are the measurements. The solid green curve represents the simulation, which is decomposed into four components: CXB, the source itself, bright X-ray sources entering the FOV around the source, and charged particles. The bottom panel shows the fit residuals.}
\label{fig:spectrum_fit}
\end{figure*}

\begin{figure}
\centering
\includegraphics[scale=0.4]{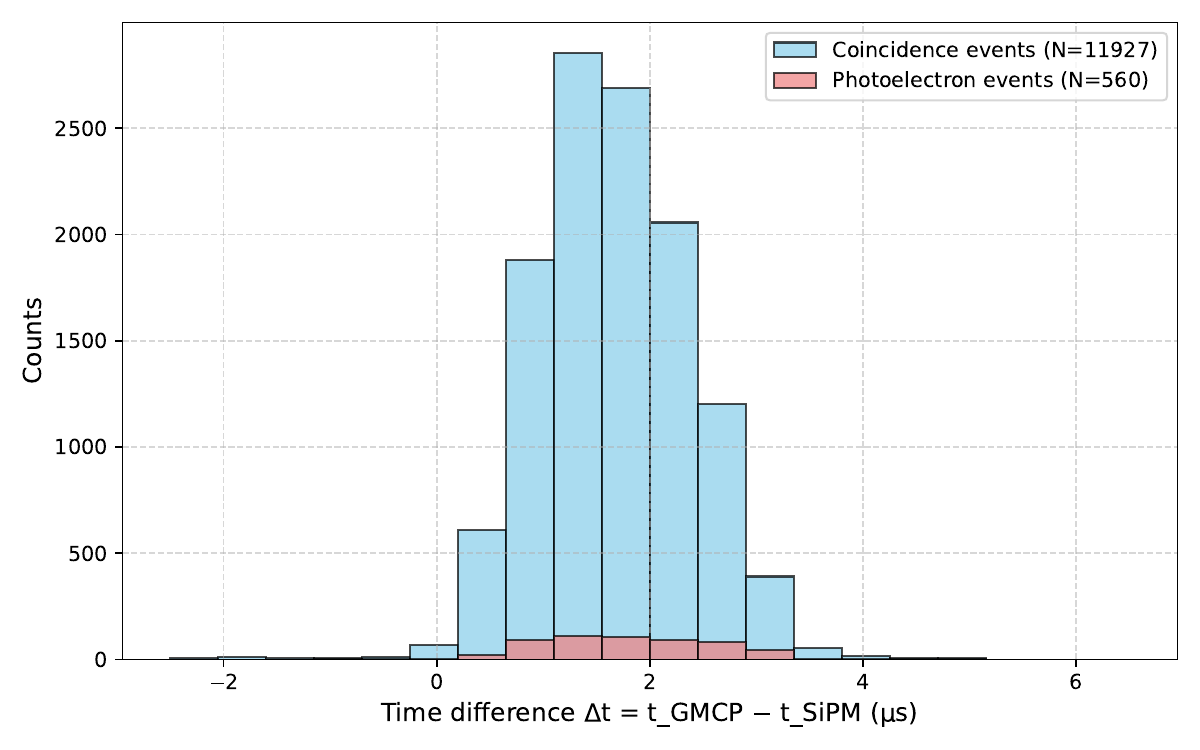}
\caption{ Distribution of the time difference $\Delta t = t_{\rm GMCP}-t_{\rm SiPM}$ for coincidence events and photoelectron events. The concentration of events within the $\pm 3~\mu\mathrm{s}$ window demonstrates the effectiveness of the time-coincidence identification.
}
\label{fig:coincidence_dt}
\end{figure}

Figure \ref{fig:adc_cir} shows the results of distinguishing between charged particles and photoelectrons based on the relationship between circularity \cite{kitaguchi2018optimized} and energy. The left panel shows the results obtained from ground-based simulations, where the background rejection rate after the linear cut is 94.06\% and the photon retention rate is 98.81\%, while the right panel presents the in-orbit results.
The effects of distinguishing charged particles from photoelectrons using on-orbit measurement data are shown in Figure \ref{fig:distinguishing}, where the count rates of photons and charged particles vary with the position projection of the detector on the Earth's surface. It is clearly visible from the figure that the photons count rate does not change significantly with variations in geomagnetic latitude, while the count rate of charged particles shows obvious modulation with geomagnetic latitude. By combining the count rate of cosmic rays at low latitudes with the simulation results, it can be concluded that the fraction of photoelectrons misidentified as cosmic rays is less than 5\%. In addition, the time-coincidence analysis among the GMCP timestamps, Topmetal signals, and the anti-coincidence detector confirms that, for the observations of Sco~X-1 and Swift~J1727.8$-$1613, the events identified as photoelectrons by the algorithm account for $(4.70 \pm 0.19)\%$ of all coincidence events. The distribution of the time differences $\Delta t = t_{\rm GMCP}-t_{\rm SiPM}$ for both coincidence-tagged events and algorithm-identified photoelectron events is shown in Fig.~\ref{fig:coincidence_dt}, illustrating the clustering of true coincident events within the $\pm3~\mu\mathrm{s}$ time window. Considering that some cosmic-ray events depositing energy in the detector gas may pass through the collimator from above or below and are therefore more difficult to discriminate, and noting that our 30.2° field of view corresponds to roughly 1/30 of the full sky, these events constitute only a small fraction of the total. Consequently, the overall fraction of cosmic rays misidentified as photoelectrons is below 10\%, which is in good agreement with the ground-based simulation results. The consistency between these two approaches allows for an estimation of both photon retention and cosmic-ray rejection efficiencies, thereby validating the accuracy of the track discrimination algorithm developed on the ground.
An example of the track images obtained after the photons distinction from the in-orbit measurements is shown in Figure \ref{fig:track_image}, where the first row shows photoelectron events and the second row shows high-energy charged particle events.

In contrast to the narrow-field, imaging background-rejection strategy adopted by IXPE—which relies on spatial selections and morphology-based filters to suppress instrumental background in the focal-plane detectors—our wide–field-of-view instrument is predominantly affected by photon background. Benefiting from the track-parameter–based selection developed in this work, more than 90\% of the photon background can be efficiently removed. The resulting photon misidentification rate is at the level of 1\%–2\%, comparable to the performance reported for IXPE \cite{di2023handling, di2022weighted}. Despite the fundamentally different background environments and rejection methodologies, both approaches achieve similarly low residual cosmic-ray background levels, ensuring that the remaining contamination has only a minimal impact on the derived polarization measurements.

% Fig.\ref{fig:track_image} shows some typical track images measured in orbit: the first row shows photoelectron tracks, and the second row shows charged particle tracks. 

\begin{figure*}
\centering
\includegraphics[scale=0.43]{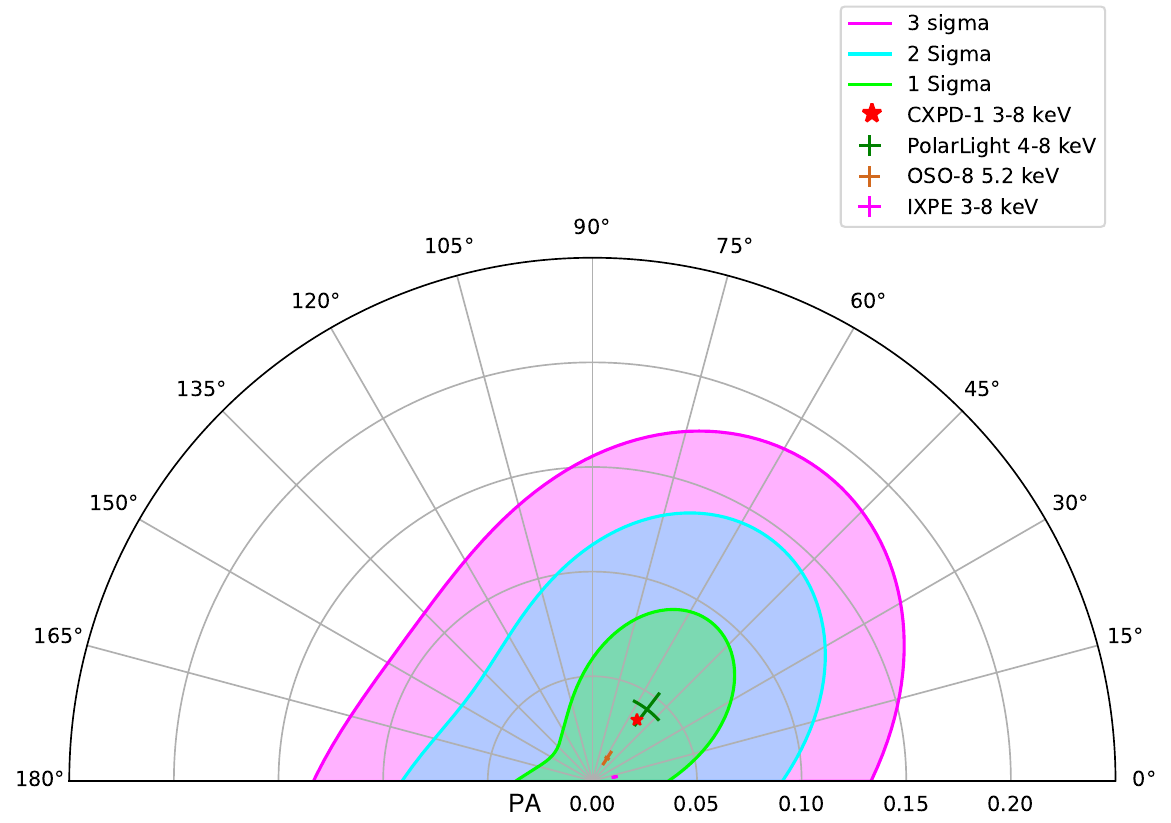}
\includegraphics[scale=0.43]{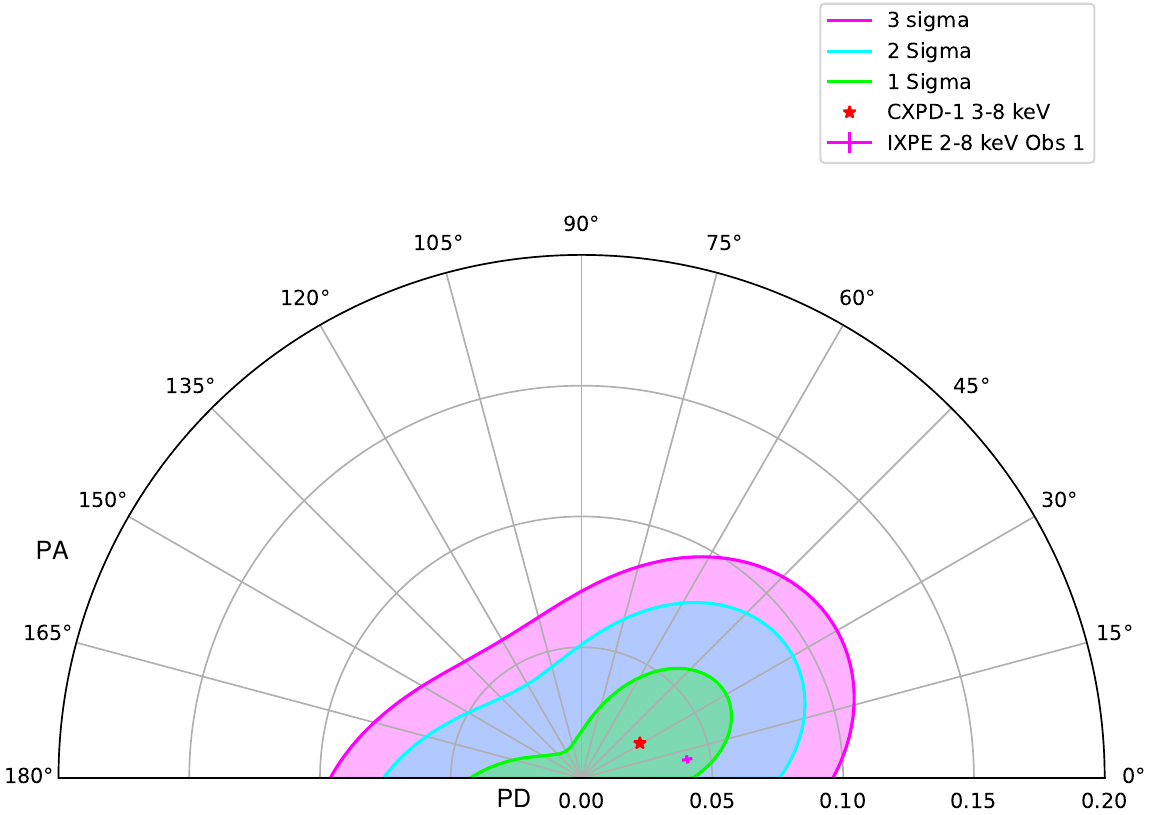}
\caption{Left: Polarization Observation Results of Sco X-1 in the 3-8 keV Energy Range. Right: Polarization Observation Results of Swift J1727.8-1613 in the 3-8 keV Energy Range.}
\label{fig:polarization}
\end{figure*}
\begin{figure*}
\centering
\includegraphics[scale=0.4]{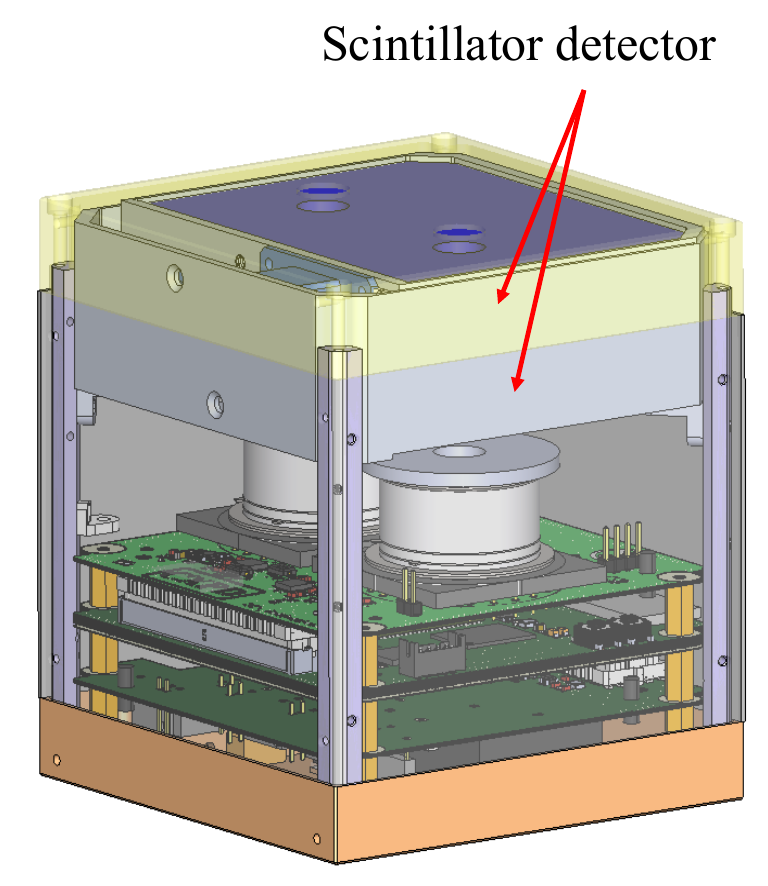}
\includegraphics[scale=0.12]{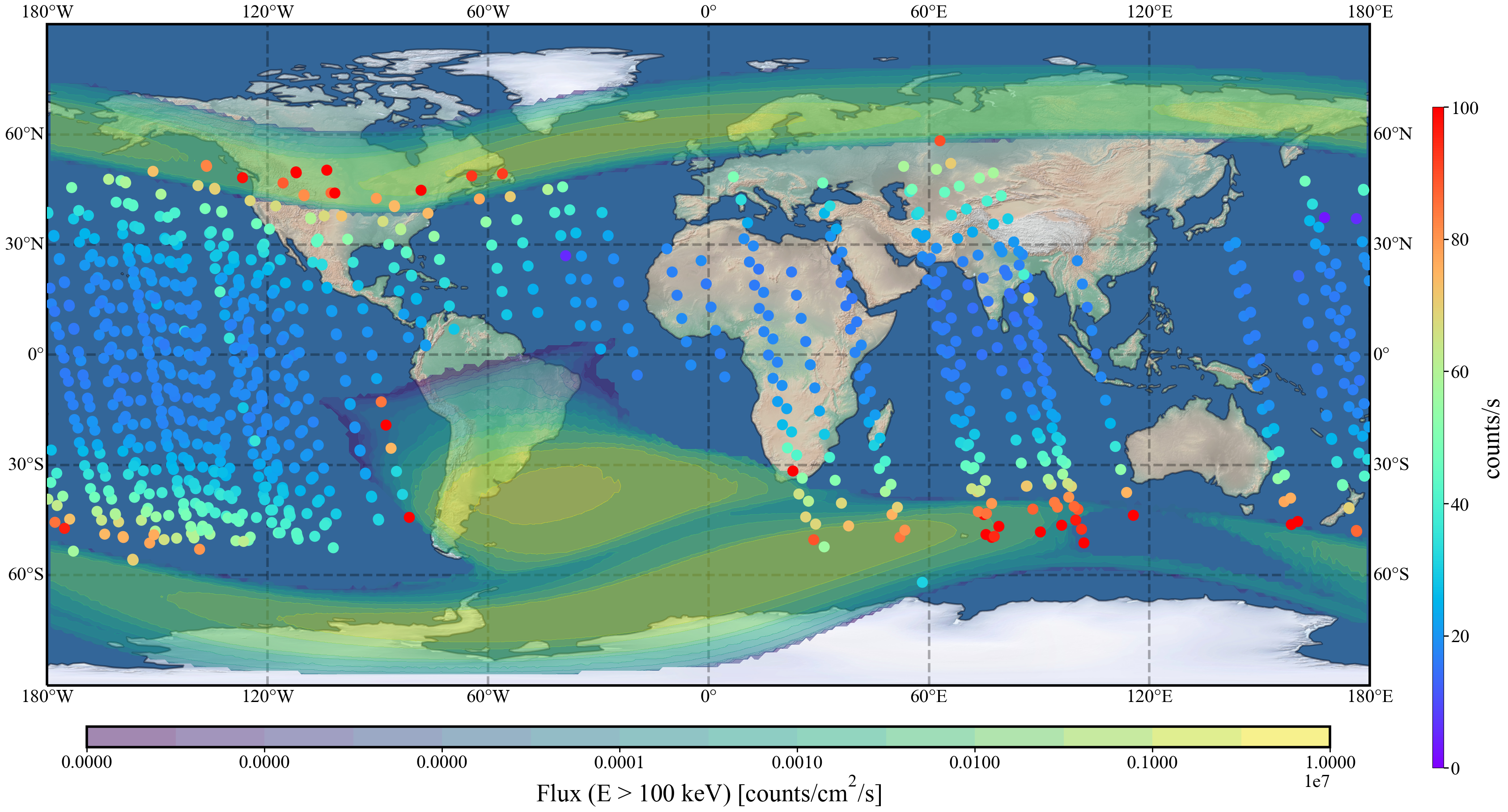}
\caption{Left: Schematic drawing of the Anticoincidence detector. 
Right: The count rate of the anticoincidence detector (points) overlaid on the flux map (contours) of trapped high-energy (\textgreater 100 keV) electrons in the orbital plane of CXPD. 
The points represent the charged particle count rates recorded at one-second intervals.
}

\label{fig:anti}
\end{figure*}

\subsection{Spectrum analysis}
\label{sec: Spectrum analysis}
To evaluate the contributions of different components to the measured energy spectra, we simulated the detector response using the Star-XP software framework \cite{yi2024star}. Star-XP incorporates Geant4-based particle–matter interactions to compute the energy deposited in the detector gas and includes a detailed digitization chain that generates realistic simulated tracks. During the digitization process \cite{AGOSTINELLI2003250}, we applied a gain variation sampled from a uniform distribution to each event to account for the small temperature-induced gain fluctuations.

In the simulation, we included all relevant spectral components: the target X-ray source, nearby bright X-ray sources, the cosmic diffuse X-ray background (CXB), and the background induced by high-energy charged particles. For Sco X-1, we adopted the spectral model and parameters derived from the IXPE observation paper \cite{la2024highly}. Although the source state and corresponding spectra of Sco X-1 can vary rapidly along its Z track, the spectral shape in the 2–8 keV range remains similar among different states, and the adopted model provides a representative description for the energy band relevant to our detector \cite{long2022significant, titarchuk2014x}. For Swift J1727.8-1613, which was observed between August 28 and September 6, i.e., prior to the IXPE campaign \cite{ingram2024tracking}, we use a power-law spectrum consistent with the hard-to-hard-intermediate state reported in the early phase of the outburst, with the photon index determined based on the 2–4 keV and 4–10 keV fluxes provided by MAXI. For other bright X-ray sources that entered the FOV, we used the models from \cite{feng2023orbit, dagoneau2021onboard}, as their fluxes are several orders of magnitude lower than those of the targets. The background components also follow the spectral prescriptions of \cite{feng2023orbit}.

The input exposure time of each simulation was matched to that of the observational data. In addition, the flux of the charged-particle background was corrected to account for the dependence on geomagnetic latitude, following the results of \cite{huang2021modeling}, which show that although the overall flux varies with latitude, the spectral shape remains largely unchanged.

The simulated spectra are compared with the measured spectra (without background rejection) in Fig.\ref{fig:spectrum_fit}. The left and right panels correspond to Sco X-1 and Swift J1727.8-1613, respectively. The results indicate that within the 30.2° FOV of the CubeSat, the contributions from the CXB and nearby bright sources are negligible compared with the brightness of the two target sources. In contrast, high-energy charged particles provide a more substantial contribution, which can, however, be reduced by an order of magnitude using the background rejection algorithm described in the previous section.

\subsection{Polarization analysis}
We conducted data selection and reconstruction for Sco X-1 and the transient source Swift J1727.8-1613. According to the spectral fitting results described in \ref{sec: Spectrum analysis}, after removing the charged-particle background using the background-rejection algorithm described in \ref{sec: Observations and Data Reduction}, the residual detector background was much lower than the source count rate. Before reconstructing the photoelectron azimuthal angles \cite{huang2021simulation}, we applied a clustering algorithm to identify charge clusters in each event and excluded edge events by discarding those with non-zero signals in the boundary pixels. 

The study and correction of the system effects of the detector indicated that the residual modulation degree of CXPD was less than 1\% in the energy range of 3-8 keV, which can be neglected relative to the statistical error of the observed photon counts. Building on previous studies, we assume the background to be unpolarized \cite{feng2020re}. For Sco X-1, a total of 3502 events in the 3–8 keV band were used for polarization analysis, among which approximately 89 events were estimated to be background according to the simulation results. 
For Swift J1727.8$-$1613, 8649 events in the same energy band were used, with an estimated background contribution of 353 events. 
The band-averaged modulation factors were derived by weighting the calibration data with the measured source spectra in each energy interval. Based on these results, the minimum detectable polarization (MDP) values were found to be 16.9\% for Sco X-1 and 10.7\% for Swift J1727.8$-$1613.

Using the Stokes parameters, we derived the polarization results of Sco X-1 and Swift J1727.8–1613 in the 3–8 keV energy band, as shown in Fig. \ref{fig:polarization}. Because of the limited number of detected events, no significant polarization is observed for either source within statistical uncertainties. For Sco X-1, the measured polarization degree has a central value of 3.6\%, as shown in the left panel of Fig. \ref{fig:polarization}. For comparison, we also include previous measurements from OSO-8, PolarLight, and IXPE, which are broadly consistent with our result within the uncertainties. For Swift J1727.8–1613, the polarization degree is found to have a central value of 2.6\%, and the first IXPE observation of this source is also shown for reference. Given the limited observation time and effective area of CXPD, possible polarization evolution cannot be constrained within the current error bars. Future coordinated observations with instruments of larger effective areas, together with additional CXPD cubic detectors and LPD units, are expected to provide more precise polarization measurements of transient sources.

\subsection{Charged particles observed by Scintillator detector}
The left panel of Fig.\ref{fig:anti} shows the schematic position of the two-layer Scintillator detectors in the CXPD CubeSat. As noted earlier, the scintillator detectors can only record the time information of detected events, which nevertheless allows us to monitor how the count rate of high-energy charged particles varies along the CubeSat orbit and thus to verify the charged-particle anomaly belt model.

The right panel shows the projection of the anticoincidence detector count rates onto the orbital plane of CXPD. The count rate of the anticoincidence detector (points) is overlaid on the flux map (contours) of trapped high-energy (\textgreater 100 keV) electrons in the orbital plane of CXPD. The points represent the count rates recorded at one-second intervals.

From the results shown in the figure, it can be seen that the count rate significantly increases when the CubeSat approaches the charged particle anomaly region, further confirming the validity of the on-off switching strategy implemented for the gas detector GMPD.

\section{Discussion and conclusion}\label{section5}
The in-orbit operation of the CXPD CubeSat provided important insights into the performance of the payload in space. It validated the encapsulation and hermeticity of the detectors and assessed the feasibility of GMCP-based detectors for space applications. In addition, the stability and radiation resistance of the GMCP and Topmetal chips were evaluated under the high-background environment of space. This mission also represented the first time that an anti-coincidence detector has been used for an X-ray polarimeter. By performing time coincidence between the events recorded by the anti-coincidence detector and those recorded by the gas detector, the accuracy and feasibility of the background rejection algorithm were directly verified.

Ground-based calibration experiments, including energy calibration, gain variation with environmental temperature, and polarization detection performance, were performed prior to launch. In orbit, observations of two bright X-ray sources, the continuously bright Sco X-1 and the bursting X-ray binary Swift J1727.8-1613, allowed verification of the simulation framework \cite{yi2024star}, background model, and background rejection algorithm developed on the ground. The polarization measurements for these two sources did not yield statistically significant results due to the limited effective area. Nevertheless, the measured values are consistent with those obtained by IXPE and PolarLight within the uncertainties. Future observations with the subsequent CubeSats, which have larger effective areas, will allow calibration measurements of standard sources in space, such as the Crab.

CXPD represents the first CubeSat from the team and features a wide FOV of 30.2°, providing a technical and space-environment assessment for the planned 90° wide-FOV design of the POLAR-2/LPD payload on the China Space Station. The experience gained from detector assembly, observation planning, in-orbit attitude control, electronics firmware update, and scientific data processing workflow laid the foundation for subsequent CubeSats. Building on this foundation, CXPD No.02–04 were launched in May 2025, incorporating the Tolmetal-L chip intended for future LPD applications, with an effective area roughly ten times larger than Topmetal-II, smaller pixels, and a wider FOV. These CubeSats aim to further test the wide-FoV LPD design, including off-axis polarization reconstruction algorithms \cite{feng2025polarization}, and are expected to facilitate more extensive scientific observations due to their larger effective area and wider FOV.

\section*{Acknowledgements}
This work is supported by the National Natural Science Foundation of China (Grant Nos. 12027803, U1731239, 12133003), the National Key R\&D Program of China (Grant Nos. 2024YFA1611700, 2023YFE0117200), the Innovation Project of Guangxi Graduate Education (No. YCBZ2025045), and the Guangxi Talent Program (“Highland of Innovation Talents”).

\bibliographystyle{apj}
\bibliography{ref.bib}

\end{document}